\documentclass[twocolumn]{aastex702}

\newcommand{\jk}{(J-K_s)_0}
\newcommand{\mj}{M_J}

\begin{document}

\title{CHASE Spectral Survey I: Measuring Contamination in the Color-selected J-region Asymptotic Giant Branch Standard Candle}

\author[0000-0002-8623-1082]{Siyang Li}
\affiliation{Department of Astronomy, University of California, Berkeley, CA 94720-3411, USA}
\email{seanli@berkeley.edu}

\author[0000-0003-3460-0103]{Alexei V. Filippenko}
\affiliation{Department of Astronomy, University of California, Berkeley, CA 94720-3411, USA}
\email{afilippenko@berkeley.edu}

\author[0000-0002-7894-1463]{Arief Ahmad}
\affiliation{Department of Astronomy, School of Science, The University of Tokyo, 7-3-1 Hongo, Bunkyo-ku, Tokyo 113-0033, Japan}
\affiliation{National Astronomical Observatory of Japan, 2-21-1 Osawa, Mitaka, Tokyo 181-8588, Japan}
\email{ariefahmad@g.ecc.u-tokyo.ac.jp}

\author[0000-0002-6124-1196]{Adam G. Riess}
\affiliation{Space Telescope Science Institute, 3700 San Martin Drive, Baltimore, MD 21218, USA}
\affiliation{Department of Physics \& Astronomy, Johns Hopkins University, Baltimore, MD 21218, USA}
\email{ariess@stsci.edu}

\author{Stefano Casertano}
\affiliation{Space Telescope Science Institute, 3700 San Martin Drive, Baltimore, MD 21218, USA}
\email{stefano@stsci.edu}

\author[0000-0001-5955-2502]{Thomas G. Brink}
\affiliation{Department of Astronomy, University of California, Berkeley, CA 94720-3411, USA}
\email{tgbrink@berkeley.edu}

\author[0000-0001-8089-4419]{Richard Anderson}
\affiliation{Institute for Astrophysics and Geophysics, Georg-August-Universit{\"a}t G{\"o}ttingen, Friedrich-Hund-Platz 1, 37077 G{\"o}ttingen, Germany}
\email{richard.anderson@uni-goettingen.de}

\author[orcid=0000-0001-6169-8586,gname='Caroline',sname='Huang']{Caroline D. Huang}
\affiliation{Center for Astrophysics | Harvard \& Smithsonian, 60 Garden St., Cambridge, MA, 02138, USA}
\email{caroline.huang@cfa.harvard.edu}

\author[0000-0003-3889-7709]{Louise Breuval}
\affiliation{European Space Agency (ESA), ESA Office, Space Telescope Science Institute, 3700 San Martin Drive, Baltimore, MD 21218, USA}
\email{lbreuval@stsci.edu}

\author[0009-0003-6827-6396]{Noriyuki Matsunaga}
\affiliation{Department of Astronomy, School of Science, The University of Tokyo, 7-3-1 Hongo, Bunkyo-ku, Tokyo 113-0033, Japan}
\email{matsunaga@astron.s.u-tokyo.ac.jp}

\author[0009-0004-9450-2489]{Gracelynn Jost}
\affiliation{Department of Astronomy, University of California, Berkeley, CA 94720-3411, USA}
\email{gracelynnjost@berkeley.edu}

\author[0009-0001-1604-4118]{Yash Mehta}
\affiliation{Department of Astronomy, University of California, Berkeley, CA 94720-3411, USA}
\email{yash_mehta@berkeley.edu}

\author[0009-0002-8370-9551]{Elaina Sadler}
\affiliation{Department of Astronomy, University of California, Berkeley, CA 94720-3411, USA}
\email{elaina_sadler@berkeley.edu}

\author[0009-0008-7952-5702]{Julian Shapiro}
\affiliation{Department of Astronomy, University of California, Berkeley, CA 94720-3411, USA}
\email{julianshapiro@berkeley.edu}

\begin{abstract}
The J-region asymptotic giant branch (JAGB) method for distance measurements uses the near uniform luminosity of carbon-rich AGB stars as a standard candle. Its application rests on an assumed but rarely examined premise: the stars contained in the JAGB color--magnitude selection box are primarily carbon-rich AGB stars. We investigate the purity of the photometric selection in the Milky Way with an optical spectroscopic survey ($\sim3590$--$7150$\,\AA) called CHASE (Carbon stars, Hubble tension, And Spectroscopic Exploration of JAGB populations)
using the Shane 3\,m telescope at Lick Observatory. These stars are classified using molecular bandhead step indices, which recover 100\% of known classifications. We report the observations of 76 stars (54 within the standard JAGB selection box of $1.5<\jk<2.0$ mag) and find only 13 of 54 stars are carbon-rich, a $76\pm6\%$ oxygen-rich contamination under blind photometric selection. This
contamination remains high ($71\%$) even within $\pm0.4$\,mag of the JAGB peak
itself, showing it is not confined to
the edges of the selection box.
This inhomogeneity may explain why recent literature find field-to-field and methodological variations of $0.1$--$0.2$\,mag that now dominate its systematic error budget. These variations could be caused, in part, by inhomogeneous stellar populations that consist of differing spectral types, and hence differing mean magnitudes.
\end{abstract}

\keywords{Carbon stars --- Asymptotic giant branch stars --- Distance indicators
--- Stellar spectroscopy}

\section{Introduction}
The Hubble constant, $H_0$, measured from early-Universe physics disagrees with
the value measured from the local distance ladder at the $6.4\sigma$ level:
Planck cosmic microwave background data under Lambda Cold Dark Matter ($\Lambda$CDM) imply
$H_0=67.4\pm0.5\,\mathrm{km\,s^{-1}\,Mpc^{-1}}$ \citep{Planck2020}, while the
covariance-weighted consensus of local distance-ladder indicators from the
Local Distance Network gives
$H_0=73.50\pm0.81\,\mathrm{km\,s^{-1}\,Mpc^{-1}}$ \citep{Casertano2026}. This difference is named the Hubble tension. Determining whether this tension reflects new physics or
an unrecognized systematic requires
independent routes to $H_0$ to probe the distance-ladder. The J-region asymptotic giant branch (JAGB) method
\citep{WeinbergNikolaev2001, MadoreFreedman2020, Ripoche2020}, a
near-infrared distance indicator built on carbon-rich AGB stars, is one such route, and has recently been extended to James Webb Space Telescope (JWST) distances
in Type Ia supernova host galaxies as part of an emerging JAGB-based
determination of $H_0$ \citep{LeeJWST2024, LeeJWSTH02024, Li2024, Li2025}. Its reliability as an $H_0$
anchor rests on a routinely repeated but rarely examined premise: the
JAGB color--magnitude selection box is dominated by carbon-rich stars. This
premise has been asserted based on photometric arguments\footnote{The JAGB
color--magnitude selection traces the red near-infrared sequence
first isolated in 2MASS color--magnitude diagrams of the Magellanic Clouds by
\citet{NikolaevWeinberg2000}, who termed it ``Region J.''
\citet{MadoreFreedman2020} adopted this photometric definition for the JAGB
method, noting that JAGB stars ``almost certainly contain a subset of the
spectroscopically classified carbon-rich population,'' but explicitly designed
the method to avoid relying on spectral or time-domain confirmation. M\&F 2020
adopted the photometric Region~J definition without any new spectroscopic
confirmation.}, and no published
work has directly tested and characterized it spectroscopically.

In the 1980s, carbon stars were proposed as extragalactic distance indicators: \citet{Richer1981} characterized the optical
luminosity function (LF) of carbon stars in the Large Magellanic Cloud (LMC), and a sequence of $I$-band
surveys extended that measurement to NGC~205, M31, NGC~300, and NGC~55
\citep{Richer1984, Richer1985a, Richer1985b, Pritchet1987}, alongside a
broader narrow-band photometric survey of carbon and M-type stars across five
Local Group galaxies \citep{Cook1986}. These 1980s studies already remarked
on the small scatter and apparent galaxy-to-galaxy constancy of the
carbon-star mean magnitude, a property later reviewed explicitly as a
standard-candle diagnostic by \citet{Battinelli2005}. The JAGB method builds
directly on this optical-era result, moving the measurement into the
$J$, $H$, and $K$ bands, where carbon-star luminosities are both brighter and
less dispersed.

\citet{MadoreFreedman2020}, \citet{FreedmanMadore2020}, and \citet{Ripoche2020} formalized this into
the modern JAGB method, exploiting the observation that carbon-rich AGB stars
occupy a narrow range of $J$-band luminosity with little color dependence.
\citet{Lee2021} calibrated a Milky Way zero point of
$\mj = -6.14\pm0.05\,(\mathrm{stat})\pm0.11\,(\mathrm{sys})$ from 153 JAGB
stars drawn from existing Milky Way carbon-star catalogs and a color cut. The method is attractive as an
independent rung of the distance ladder: it needs only single-epoch
near-infrared photometry, which suffers less from interstellar extinction
than optical bands, and takes advantage of the NIR capabilities of JWST to
reach far distances. Its calibration
and standard-candle behavior have since been examined further in the Milky
Way and Local Group \citep{Parada2021, Parada2023, Madore2022},
applied to individual nearby galaxies in the same series such as WLM \citep{LeeWLM2021} and M33
\citep{LeeM332022}, and extended to larger ground-based multigalaxy samples
through the Araucaria Project and Magellan imaging
\citep{Zgirski2021, LeeMagellan2024}.

The JAGB method rests on the premise that carbon-rich AGB stars occupy a narrow, near-constant near-infrared magnitude \citep{MadoreFreedman2020}; recent work, however, has suggested that further standardization for the JAGB method may be warranted.
\citet{Li2024} found intrahost JAGB brightness differences of up to
$\sim0.2$\,mag that significantly exceed statistical uncertainties, together
with nonuniform LF shapes. \citet{Li2025} measured a
difference of $0.11\pm0.022$\,mag between two NGC~4258 calibrator fields, which
dominates their $H_0$ systematics. \citet{Parada2021,Parada2023} found different
JAGB median brightness and LF shape between the LMC and Small Magellanic Cloud (SMC), and \citet{Magnus2024} propose that the JAGB zero
point is not independent of metallicity and that oxygen-rich contamination of
the purely color-selected box matters. These discrepancies raise the question of what causes them.

By construction, JAGB distances are inferred from photometry alone. If the JAGB box contains a
mixture of stellar types with different luminosities, that mixture is a natural
explanation for the observed LF asymmetry. It is also a natural explanation
for field-to-field variation, specifically if the relative fractions or
luminosity distributions of those types themselves vary from field to
field. Testing either requires spectroscopy, and existing spectra cannot
fill that gap.
Existing carbon-star and long-period-variable catalogs are also not built for
this purpose: catalogs such as those of \citet{ChenYang2012},
\citet{Li2024LAMOST}, \citet{Abia2022}, and the Gaia DR3
long-period-variable catalog \citep{GaiaDR3} select stars as carbon-rich or
as long-period variables in general, and only incidentally fall inside the
JAGB color box (Figure~\ref{fig:selection}). The survey presented here, CHASE (Carbon stars, Hubble tension, And SPectroscopic Exploration of JAGB populations), by contrast, selects
targets directly by J-region membership, so its census of the box is direct
rather than inferred from a broader population that happens to overlap it.
This also inverts the approach used to build the method's own Milky Way zero
point: \citet{Lee2021} began from known carbon-star catalogs and checked
which members fell inside the color box, whereas the present survey, CHASE begins from the box
itself and asks what occupies it.
We cross-matched the JAGB selection presented below against a broad set of accessible
libraries of Milky Way stellar spectra spanning optical through near-infrared
wavelengths, since AGB stars peak in flux at $1$--$2\,\mu$m (Table~\ref{tab:catalogs}). MILES \citep{Sanchez-Blazquez2006},
the X-Shooter Spectral Library DR1 \citep{Gonneau2016}, RAVE \citep{Steinmetz2006},
GALAH \citep{DeSilva2015}, SDSS \citep{York2000}, the extended IRAS LRS atlas
\citep{Sloan2025}, and the Palomar Gattini-IR variable census \citep{Earley2025}
return \emph{zero} matches to a bright
($G\le13$ mag) sample: these surveys either avoid the brightest stars, cover
the southern hemisphere, or simply do not overlap our specific target list. LAMOST \citep{Cui2012}, which is faint-star oriented, matched 19 stars.
The X-Shooter Spectral Library DR3 \citep{Verro2022} matched 1 star, and APOGEE DR17
\citep{Abdurrouf2022}, a near-infrared $H$-band survey, matched 4 stars (3 with a
retrievable spectrum). The near-infrared IRTF library had 2 matches.
The principal optical carbon-star atlas \citep{Barnbaum1996} has archived
spectra for only 39 stars, of which 4 overlap our sample. Existing
classifications are also heterogeneous, spanning decades of photographic and
digital work with incompatible notations, and we find direct disagreements
between sources (e.g.,\ RY\,Dra is C--N3III in SIMBAD but C--J4 in
\citealt{Barnbaum1996}). No existing dataset can characterize the JAGB box
uniformly. We therefore obtained the first set of homogeneous, flux-calibrated, telluric-corrected optical
spectra of a JAGB-selected Milky Way sample as a part of the CHASE survey. Because every spectrum passes through a single, uniform reduction and
classification pipeline, cross-catalog calibration differences and the
heterogeneous classification conventions noted above cannot be the origin of
any oxygen-rich contamination identified within the box;
Section~\ref{sec:jagbcomp} confirms this directly, finding zero
disagreements with the literature among the 46 box stars with an existing
classification.

The remainder of this paper is organized as follows. The target selection,
observations, spectral reduction, and photometry and reddening are described
in Section~\ref{sec:data}. We present our bandhead-based spectral classification in Section~\ref{sec:results}  and its resulting composition of the
JAGB box. Section~\ref{sec:discussion} discusses the implications of this
inhomogeneity, its sensitivity to the color cut, its reconciliation with
the Milky Way contamination estimate of \citet{Magnus2024}, how far it generalizes beyond the solar neighborhood, and
the caveats of our analysis. We summarize our conclusions in
Section~\ref{sec:conclusions}.

\section{Data}
\label{sec:data}
CHASE is designed to test whether the JAGB selection box is dominated by
carbon-rich stars, using a well-defined Milky Way parent sample
(Section~\ref{sec:targetsel}) and an optical spectroscopic
setup. The configuration used throughout this survey
(Section~\ref{sec:obs}) was chosen so its combined blue+red wavelength
coverage, $\sim3590$--$7150$\,\AA, spans all nine molecular bandheads used
for our bandhead-step classification (Section~\ref{sec:bandhead}): six TiO
heads and three C$_2$ Swan heads, the diagnostic bands that separate
carbon- from oxygen-rich chemistry. Of the 357,002-star parent sample, 76
stars have been observed to date, of which 72 satisfy the parent selection
cuts and 54 fall in the JAGB color-magnitude box; this paper reports the
spectroscopic classification of these 76 stars, finding 13 of the 54 box
stars are carbon-rich, and the resulting composition of the JAGB box. CHASE
is an ongoing survey, and additional observing seasons are planned to
expand this sample. The data analyzed in this paper will be made publicly available upon publication.

\subsection{Target Selection}
\label{sec:targetsel}
The parent sample is drawn from a Gaia DR3 \citep{GaiaDR3} query requiring
Galactic latitude $b\ge10^{\circ}$, $G$-band magnitude $G\le13$ mag,
color $\mathrm{BP}-\mathrm{RP}\ge1.3$ mag, and
parallax signal-to-noise $\varpi/\sigma_{\varpi}>10$, cross-matched to the Two Micron All Sky Survey \citep[2MASS;][]{TwoMASS}. The JAGB selection box is constructed by applying cuts of
$1.5 <\jk<2.0$ mag and $-9.14<\mj<-3.14$ mag (i.e.,\ symmetric, $\pm3$\,mag around the JAGB from 
\citealt{Lee2021}), where $\jk$ and $\mj$ are, respectively, dereddened color
and Bailer-Jones-distance-based absolute magnitude as described in
Section~\ref{sec:phot}. We adopt a wide magnitude range to define the initial selection to investigate stars near the horizontal (magnitude) edges of the JAGB box and their potential to scatter into and out of the JAGB box; Section~\ref{sec:colorcut} separately investigates sensitivity to the box's color (vertical) edges. Targets were scheduled by an earliest-deadline-first
algorithm at airmass $\le2$, a limit to avoid the increased atmospheric extinction and reduced signal-to-noise of observations at higher airmass. Each star was individually pre-selected from
this parent catalog based on that night's visibility, and stars were
observed one at a time following this schedule.

Of the 76
unique science stars observed based on scheduling constraints at Lick Observatory, 72 satisfy the parent selection and 54 fall in
the JAGB selection box under the corrected photometry of Section~\ref{sec:phot}. These 54 constitute the primary JAGB analysis
sample; we explore the effects of color cuts on the inclusion of the remaining stars outside these 54 in the Discussion. Four stars fail the parent cuts and are excluded: M-type Miras with
$\varpi/\sigma_{\varpi}\le10$, observed as backup targets when too few
parent-sample candidates were available in their scheduling window, and
retained for the classifier-validation test of Section~\ref{sec:bandhead}. Table~\ref{tab:skypos} lists Galactic coordinates and Bailer-Jones distances
(Section~\ref{sec:phot}) for the full observed sample.

\begin{figure*}[t]
\centering
\includegraphics[width=0.92\textwidth]{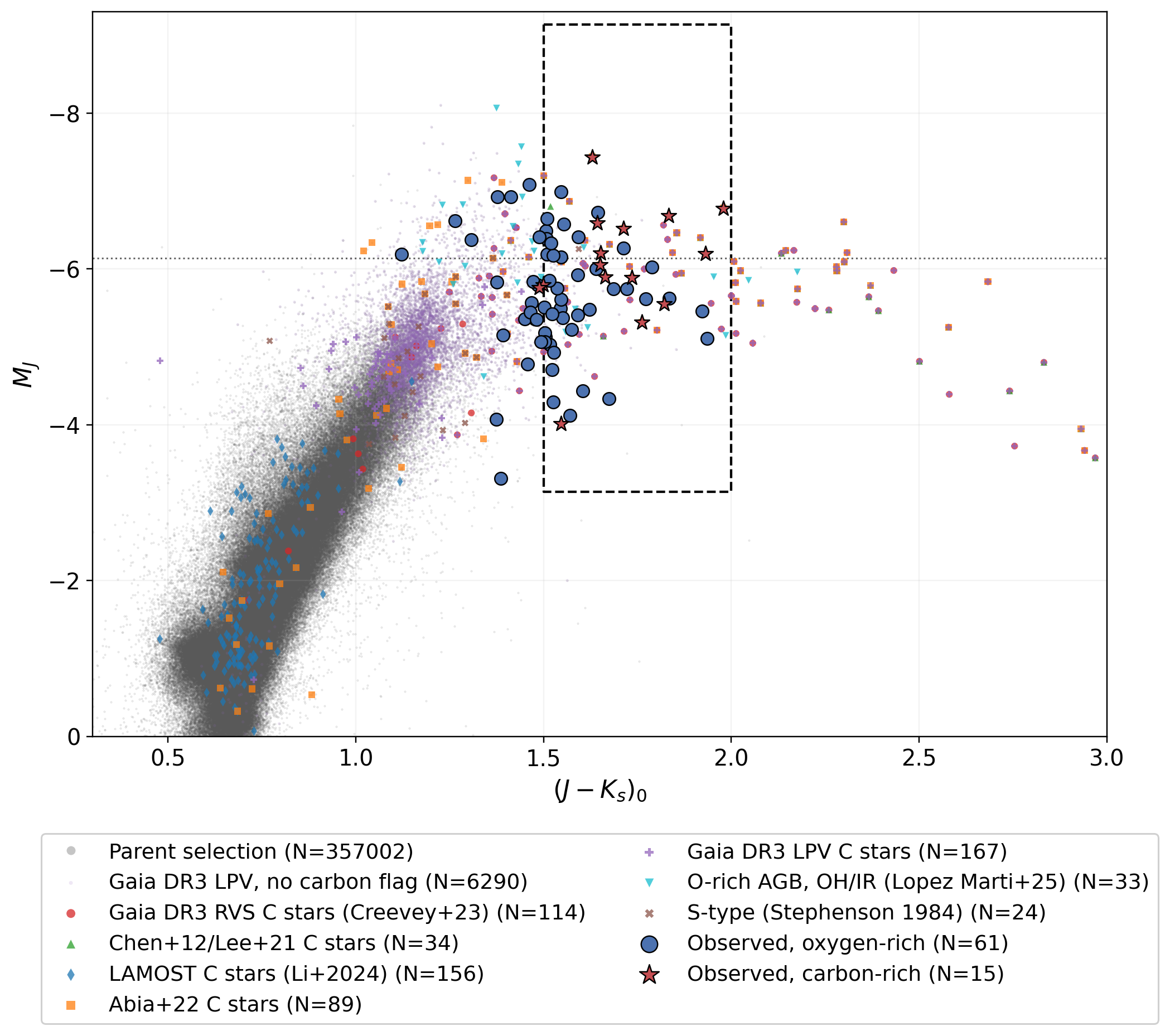}
\caption{Selection context: the full Gaia~DR3 parent-selection
population (357{,}002 stars, gray) in the dereddened
$M_J$--$\jk$ plane, with the JAGB color box (dashed) and the
\citet{Lee2021} peak (dotted horizontal line). Colored markers show literature
carbon-star catalog matches: \citet{Creevey2023} ($N=114$), \citet{Lee2021,ChenYang2012}
($N=34$), \citet{Li2024LAMOST} ($N=156$), \citet{Abia2022} ($N=89$), and the
Gaia DR3 long-period-variable catalog ($N=167$). The much larger population of
long-period variables with no carbon flag in that same catalog ($N=6290$;
not a confirmed oxygen-rich classification, since the catalog's only
classification column is a binary carbon flag) is also shown, in light purple,
for context. Also shown are two independent, non-carbon literature
catalogs: oxygen-rich AGB stars identified via OH maser emission
(\citealt{LopezMarti2025}; $N=33$) and S-type stars
(\citealt{Stephenson1984}; $N=24$). The 76 Shane/Kast
observed science stars are overplotted on top, colored by this
work's spectroscopic classification (red star = carbon-rich, blue circle =
oxygen-rich). For stars we actually observed, our spectroscopic call is
drawn on top of and takes precedence over any literature marker
underneath. Typical photometric uncertainties are $\lesssim0.03$ mag, rising
to $\sim0.2$--$0.3$ mag for 2MASS-saturated stars (Section~\ref{sec:caveats}).\label{fig:selection}}
\end{figure*}

Figure~\ref{fig:selection} places the observed sample in this broader context: of the 357{,}002 parent-selection
candidates with valid  2MASS photometry, 364 (0.10\%) are flagged as carbon stars by at
least one of five literature catalogs: \citet{Creevey2023},
\citet{Lee2021,ChenYang2012}, \citet{Li2024LAMOST}, \citet{Abia2022}, and the Gaia DR3
long-period-variable catalog \citep{GaiaDR3}. We observed
15 of these 364 (4.1\%), a fraction set by telescope-time and scheduling
constraints rather than by any astrophysical cut. We additionally
cross-matched the parent sample against two independent, non-carbon
catalogs to test whether O-rich and chemically intermediate AGB stars also
populate the box: 33 stars match the oxygen-rich OH/IR AGB catalog of
\citet{LopezMarti2025} (13 in the JAGB box), and 24 match the S-type star
catalog of \citet{Stephenson1984} (1 in the JAGB box).

Table~\ref{tab:surveycomp} compares CHASE to these five catalogs on survey
design and box overlap: each literature catalog's carbon-star census
overlaps the JAGB box only incidentally, at a rate more than an order of
magnitude below CHASE's $71\%$.

\begin{table*}[t]
\centering
\caption{Comparison with the literature carbon-star/LPV catalogs of Figure~\ref{fig:selection}.\label{tab:surveycomp}}
\footnotesize
\resizebox{\textwidth}{!}{%
\begin{tabular}{lccccc}
\hline
Survey & Wavelength & Resolution & $N$ (total) & Selection & $J$-region overlap \\
\hline
CHASE (this work) & $3590$--$7150$\,\AA & $\sim760$ (blue) / $\sim1200$--$1500$ (red) & 76 & JAGB box + parent color/mag cuts & $54/76$ ($71\%$) \\
Galactic IR carbon stars & heterogeneous & -- & 579 & Literature compilation & $10/579$ ($1.7\%$) \\
\citep{Lee2021,ChenYang2012} & & & & & \\
LAMOST DR7 \citep{Li2024LAMOST} & $\sim3700$--$9000$\,\AA & $\sim1800$ & 3546 & LAMOST targets, post hoc ID & $0/3546$ ($0.0\%$) \\
Gaia EDR3 carbon stars \citep{Abia2022} & -- & -- & 827 & Literature catalogs + Gaia astrometry & $20/827$ ($2.4\%$) \\
Gaia DR3 Golden Sample \citep{Creevey2023} & Gaia BP/RP, $330$--$1050$\,nm & $\sim50$ & 15{,}740 & C$_2$/CN in BP/RP spectra & $42/15{,}740$ ($0.3\%$) \\
Gaia DR3 LPV catalog \citep{Lebzelter2023} & Gaia RP, $\sim630$--$1050$\,nm & $\sim50$ & 546{,}468 & RP peak separation (\texttt{isCstar}) & $42/546{,}468$ ($0.01\%$) \\
\hline
\end{tabular}
}
\tablecomments{Resolution is $\lambda/\mathrm{FWHM}$. $N$ (total) is each catalog's full carbon-star sample, not our parent selection; $J$-region overlap is the number of that catalog's stars falling in the JAGB color--magnitude box (Section~\ref{sec:targetsel}) among those also passing CHASE's own parent selection cuts (Section~\ref{sec:targetsel}) and with usable Bayestar19-dereddened photometry, so the printed fraction is a lower limit on the catalog's true box overlap, not the box overlap of the full sample. CHASE's resolution is measured directly from night-sky emission-line FWHM (5577, 6300, and 6364\,\AA) in our reduced spectra. The Galactic IR carbon-star row uses \citet{Lee2021}'s own 579-star compilation, which itself draws primarily from \citet{ChenYang2012}, cross-matched to Gaia EDR3 and originally compiled from heterogeneous photographic, photometric, and spectroscopic literature with no single wavelength or resolution. \citet{Abia2022} re-analyzes existing literature carbon-star catalogs using Gaia EDR3 astrometry and photometry rather than new spectroscopy.}
\end{table*}

\subsection{Observations}
\label{sec:obs}
Observations were obtained on 2026 February 13, March 16, and May 13 UTC with the
Kast double spectrograph on the Shane 3\,m telescope at Lick Observatory (see Table~\ref{tab:log}).
A dichroic in Kast splits each star's light into
separate blue and red channels, recording them simultaneously on two detectors with
independent gratings, so every exposure yields two spectra per star, one
per arm, which must later be joined into a single continuous spectrum (Section~\ref{sec:stitch}). All
three nights used an identical configuration: 600/4310 grism (blue arm), 600/5000
grating (red arm), d57 dichroic, $2.0''$-wide slit, unbinned. The slit was
oriented at or near the parallactic angle for each exposure, minimizing
differential light losses from atmospheric dispersion
\citep{Filippenko1982}. Measured coverage is
$\sim3590$--$5710$\,\AA\ at $1.06$\,\AA\,pix$^{-1}$ (blue) and
$\sim5240$--$7150$\,\AA\ at $1.29$\,\AA\,pix$^{-1}$ (red).

\subsection{Reduction}
\label{sec:reduction}
Data were reduced with \texttt{PypeIt} v2.0.0 \citep{PypeIt}, which computes
two candidate 1D extractions per exposure: an \emph{optimal} extraction, which
weights each pixel by its expected signal-to-noise ratio (S/N) given a fitted spatial
profile of the star's light on the detector (an initial full width at half-maximum intensity (FWHM) guess of
$4.65$\,pixel, refined per exposure from the data), and a simpler
\emph{boxcar} extraction, which just sums all pixels within a fixed
$2\farcs0$ radius unweighted. The coaddition step (below) uses the optimal
extraction by default. For 2 of the 680 extracted spectra in this reduction
(both from the same night and arm: 2026 May 13, red), the local
spatial-profile fit diverged, the sigma-rejection threshold escalated past
$100\sigma$ while rejecting nearly all fit pixels, and the resulting
optimal extraction was masked; those two exposures were effectively
down-weighted in the coaddition rather than contributing a spurious optimal
extraction.

Exposures were bias-corrected, flux-calibrated with
sensitivity functions derived from flux standards, coadded, and telluric-corrected
on the red side using a model that fits the atmosphere's absorption jointly
with a polynomial approximation of the star's own continuum shape. No
synthetic stellar template was used for this fit, as reliable, high-resolution
synthetic spectral templates that reproduce the detailed molecular-band
structure of carbon and other peculiar evolved stars do not exist, and a
mismatched template would corrupt the very bands we intend to measure. This
is a different requirement than the broadband synthetic photometry from
hydrostatic atmosphere models used in Section~\ref{sec:comarcs} to place
carbon- and oxygen-rich models in the JAGB color--magnitude plane, which does
not depend on reproducing fine spectral-line detail. Each science exposure was
flux-calibrated against the standard star minimizing
$0.5\,|\Delta X|/1.5 + 0.5\,(\Delta t/8\,\mathrm{hr})$, where $\Delta X$ is
the airmass difference and $\Delta t$ the time between the science and
standard exposures i.e.,\ the nearest standard in airmass and elapsed time,
weighted equally after normalizing to an airmass of $1.5$  and an $8$\,hr scale.
All three nights were reduced identically with this configuration.

\subsection{Blue + Red Stitching}
\label{sec:stitch}
The blue and red arms were combined into a single spectrum per star before any
further analysis. The blue arm is trimmed to $3600$--$5550$\,\AA\ (throughput
falls sharply redward of this) and the red arm to $5500$--$7200$\,\AA, a
ceiling chosen above the reddest coverage reached by any single exposure
($7180$\,\AA; the $\sim7150$\,\AA\ value above is the mean across exposures), and so the two trimmed arms
overlap over $5500$--$5550$\,\AA. To compute a
single multiplicative scale factor,
$s = \mathrm{median}(f_{\rm red})/\mathrm{median}(f_{\rm blue})$, we instead
use a wider $5400$--$5600$\,\AA\ window near the dichroic crossover for a more
robust median: pixels with $\mathrm{S/N}>3$ are selected independently in
each arm's own trimmed coverage within that window ($5400$--$5550$\,\AA\ for
blue, $5500$--$5600$\,\AA\ for red); the blue arm's flux, and its inverse-variance array (the per-pixel
statistical weight $1/\sigma^2$ used to track and propagate flux
uncertainty), are rescaled together by $s$ and $1/s^2$, respectively, and the
two arms are concatenated and the combined array is re-sorted by wavelength,
ensuring the final spectrum has a single monotonically increasing wavelength
axis.

\subsection{Photometry, Reddening, and Distances}
\label{sec:phot}
We adopt single-epoch 2MASS $J$ and $K_s$ photometry \citep{TwoMASS} and Gaia DR3
positions and parallaxes \citep{GaiaDR3}; because these are large-amplitude
pulsating stars observed at an uncontrolled phase, single-epoch photometry
adds scatter to $\mj$, which we bound in Section~\ref{sec:caveats}. Reddening is taken from the three-dimensional (3D) Bayestar19 dust map
\citep{Green2019}, queried at each star's \cite{BailerJones2021} distance, giving a
reddening value $E$ for each star; this is converted to extinction in each
band via $A_X = R_X E$, using the same coefficients as by \citet{Ripoche2020}:
$R_J,~R_H,~R_{K_s} = 0.7927, ~0.4690, ~0.3026$.

We utilize a 3D map to avoid potential biases in the reddening corrections. The targets observed here are Milky Way
stars at median distance $\sim1$\,kpc,  \emph{inside} the dust layer. A
full-line-of-sight map, one that integrates dust along the entire
line of sight rather than stopping at each star's distance, such as the
widely used Schlegel--Finkbeiner--Davis \citep[SFD;][]{SFD1998} map as recalibrated by \cite{SchlaflyFinkbeiner2011}, would therefore systematically overcorrect them. In
practice, the difference is small for this sample (median
$\Delta\jk=+0.0001$ and $\Delta\mj=-0.099$\,mag), and none of our conclusions
depends on it.

\startlongtable
\begin{deluxetable*}{lccccc}
\tablecaption{Sky position and distance of the observed sample.\label{tab:skypos}}
\tabletypesize{\small}
\tablehead{\colhead{Star} & \colhead{RA (J2000)} & \colhead{Dec (J2000)} & \colhead{$l$ (deg)} & \colhead{$b$ (deg)} & \colhead{$d_{\rm rgeo}$ (pc)}}
\startdata
ATO\_J106.0289+18.2473 & 07 04 06.95 & +18 14 50.60 & 197.9 & 10.9 & 4915 \\
BD+09\_1709B & 07 35 46.942 & +09 35 55.97 & 350.5 & 84.9 & 2742 \\
BD+24\_2541 & 13 07 53.20 & +23 37 29.17 & 186.7 & 74.7 & 805 \\
BD+33\_2151 & 11 46 01.05 & +32 55 58.30 & 211.3 & 38.4 & 867 \\
BD+37\_3051 & 18 13 57.638 & +37 09 50.11 & 154.3 & 21.5 & 749 \\
BD+45\_2569 & 17 37 21.537 & +45 54 18.16 & 79.1 & 46.8 & 766 \\
BD+51\_1329 & 07 35 21.530 & +51 32 22.52 & 183.6 & 32.0 & 1925 \\
DO\_35517 & 16 56 05.942 & +53 25 32.02 & 122.1 & 51.1 & 945 \\
DT\_Boo & 14 21 44.600 & +43 59 47.50 & 190.6 & 49.8 & 674 \\
FM\_Dra & 13 19 51.010 & +65 11 03.60 & 161.2 & 23.4 & 719 \\
GQ\_Her & 17 45 40.966 & +18 50 37.23 & 96.9 & 58.5 & 825 \\
HD134607 & 15 08 15.140 & +57 39 25.20 & 117.1 & 60.5 & 477 \\
HD\_146251 & 16 13 05.220 & +48 05 36.49 & 126.4 & 71.6 & 568 \\
HD\_47374 & 06 40 45.108 & +49 01 40.12 & 91.1 & 67.2 & 664 \\
HD\_56219 & 07 19 36.382 & +58 26 55.61 & 209.3 & 14.2 & 877 \\
IRAS\_06458+4609 & 06 49 31.872 & +46 06 27.47 & 64.3 & 23.0 & 1454 \\
IRAS\_07207+5327 & 07 24 42.000 & +53 22 03.94 & 166.5 & 27.5 & 899 \\
IRAS\_09074+1751 & 09 10 16.66 & +17 39 22.32 & 81.1 & 38.5 & 915 \\
IRAS\_15506+6650 & 15 51 00.820 & +66 41 16.00 & 75.2 & 45.9 & 1007 \\
IRAS\_16032+4407 & 16 04 50.600 & +43 59 11.40 & 166.6 & 18.5 & 914 \\
IRAS\_16577+6045 & 16 58 23.820 & +60 40 41.10 & 158.4 & 26.5 & 945 \\
IRAS\_18243+3025 & 18 26 15.919 & +30 27 44.86 & 170.0 & 18.9 & 1155 \\
IRAS\_18427+3011 & 18 44 41.765 & +30 14 29.83 & 164.1 & 26.3 & 1402 \\
IRC+50261 & 16 59 49.720 & +52 19 04.40 & 58.4 & 18.4 & 1831 \\
IRC\_+30295 & 16 41 58.834 & +32 55 29.93 & 59.8 & 14.6 & 801 \\
IRC\_+30329 & 18 12 48.322 & +25 05 57.80 & 54.4 & 40.4 & 1073 \\
IRC\_+50252 & 16 26 15.122 & +47 49 33.60 & 52.0 & 19.2 & 1220 \\
IRC\_+50278 & 18 20 54.886 & +50 31 43.00 & 74.4 & 43.8 & 1368 \\
IRC\_+50280 & 18 27 40.546 & +49 18 33.08 & 78.8 & 25.3 & 1218 \\
IRC\_+60184 & 07 55 50.700 & +57 11 53.23 & 77.7 & 23.9 & 975 \\
LEE\_195 & 07 04 58.186 & +25 16 19.81 & 160.4 & 31.1 & 2573 \\
LEE\_204 & 16 17 08.760 & +25 51 01.48 & 191.4 & 14.0 & 2148 \\
LEE\_68 & 06 57 35.806 & +20 24 05.33 & 43.6 & 44.3 & 2368 \\
LL\_Dra & 12 29 37.320 & +68 38 07.60 & 195.2 & 10.4 & 906 \\
R\_Com & 12 04 15.120 & +18 46 55.20 & 75.8 & 42.9 & 1261 \\
SV\_Leo & 10 02 41.040 & +26 41 34.80 & 50.0 & 22.2 & 2139 \\
SW\_UMi & 14 04 57.880 & +66 20 20.60 & 199.6 & 13.6 & 569 \\
S\_LMi & 09 53 43.200 & +34 55 33.60 & 75.4 & 39.1 & 1640 \\
StM\_282 & 16 30 43.181 & +48 53 47.26 & 56.5 & 15.1 & 1077 \\
StM\_414 & 17 57 27.689 & +24 30 17.24 & 189.4 & 24.3 & 1489 \\
StM\_74 & 07 17 24.526 & +17 53 35.30 & 193.2 & 17.2 & 1197 \\
TW\_Lyr & 18 23 56.308 & +39 35 04.39 & 204.9 & 32.2 & 899 \\
TZ\_Leo & 11 23 40.080 & +16 51 07.20 & 11.0 & 53.3 & 780 \\
V\_AI\_Her & 16 53 44.794 & +48 57 02.38 & 169.0 & 20.6 & 1294 \\
V\_AP\_Lyn & 06 34 33.40 & +60 56 27.46 & 166.5 & 43.5 & 482 \\
V\_AS\_Lyr & 18 37 01.068 & +27 34 34.43 & 172.5 & 15.1 & 1383 \\
V\_AU\_Gem & 07 45 27.442 & +30 46 41.23 & 81.6 & 32.8 & 1004 \\
V\_BM\_Gem & 07 20 59.004 & +24 59 57.95 & 157.8 & 21.8 & 1214 \\
V\_DK\_Cnc & 08 38 06.715 & +20 22 50.41 & 207.3 & 36.1 & 1128 \\
V\_FV\_Boo & 15 08 25.754 & +09 36 18.36 & 176.5 & 13.8 & 940 \\
V\_Leo & 10 00 01.930 & +21 15 44.40 & 29.8 & 33.4 & 1151 \\
V\_QX\_Aur & 06 57 09.029 & +47 35 21.84 & 76.4 & 19.0 & 924 \\
V\_RR\_Her & 16 04 13.40 & +50 29 56.98 & 42.0 & 32.0 & 1269 \\
V\_RT\_Lyn & 08 14 50.63 & +37 40 11.78 & 45.7 & 40.4 & 1384 \\
V\_RT\_UMa & 09 18 24.408 & +51 24 07.45 & 41.9 & 45.1 & 1973 \\
V\_RV\_Aur & 06 34 44.628 & +42 30 12.71 & 183.5 & 12.0 & 1244 \\
V\_RY\_Dra & 12 56 25.94 & +65 59 39.66 & 29.5 & 53.5 & 398 \\
V\_R\_LMi & 09 45 34.29 & +34 30 42.77 & 72.1 & 31.6 & 289 \\
V\_R\_Lyn & 07 01 18.01 & +55 19 49.66 & 81.7 & 65.2 & 934 \\
V\_SY\_Dra & 17 33 57.818 & +53 57 37.51 & 118.1 & 51.7 & 1123 \\
V\_S\_Boo & 14 22 52.94 & +53 48 37.12 & 43.3 & 22.7 & 1595 \\
V\_S\_Lyn & 06 44 34.097 & +57 54 39.74 & 94.3 & 51.2 & 987 \\
V\_T\_Cnc & 08 56 40.130 & +19 50 56.98 & 100.9 & 41.9 & 933 \\
V\_UU\_Aur & 06 36 32.837 & +38 26 43.51 & 69.6 & 47.9 & 489 \\
V\_UV\_Her & 16 45 34.111 & +12 08 11.47 & 90.2 & 37.2 & 1189 \\
V\_UW\_UMa & 13 12 01.51 & +56 22 48.29 & 79.7 & 38.0 & 985 \\
V\_UY\_Lyr & 18 53 40.829 & +46 41 09.67 & 125.9 & 48.4 & 1988 \\
V\_V456\_Her & 17 06 47.606 & +21 06 17.71 & 248.0 & 76.3 & 1269 \\
V\_V901\_Her & 16 35 33.715 & +26 28 27.12 & 203.7 & 52.6 & 769 \\
V\_VV\_Her & 16 12 29.753 & +24 53 57.12 & 111.8 & 49.2 & 1093 \\
V\_VW\_Gem & 06 42 08.587 & +31 27 17.50 & 190.0 & 51.5 & 1179 \\
V\_WX\_Ser & 15 27 47.045 & +19 33 51.80 & 67.4 & 21.9 & 687 \\
V\_Y\_CVn & 12 45 07.82 & +45 26 25.15 & 235.2 & 67.3 & 314 \\
V\_ZZ\_CVn & 13 59 11.08 & +45 28 16.14 & 211.9 & 50.7 & 674 \\
WY\_Boo & 13 59 38.470 & +27 47 14.40 & 39.4 & 74.9 & 671 \\
W\_Leo & 10 53 37.440 & +13 42 54.00 & 233.0 & 59.4 & 1051 \\
\enddata
\tablecomments{Distances are Bailer-Jones \citep{BailerJones2021} geometric ($r_{\rm geo}$) distances, as used throughout this paper (Section~\ref{sec:phot}).}
\end{deluxetable*}

\section{Results}
\label{sec:results}
\subsection{Classification from Bandhead Steps}
\label{sec:bandhead}
The spectra of cool giant stars are shaped by broad molecular absorption bands (TiO in
oxygen-rich stars, C$_2$ and CN in carbon stars) rather than the sharp atomic
lines of hotter stars. Each band consists of many rotational-vibrational
transitions that pile up at a specific wavelength at the bandhead and then
thin out to one side of it, so the absorption is deep and sharp on one side
of the bandhead and shallow on the other. Measuring which molecule dominates,
and how strongly, is the basis for distinguishing carbon-rich stars (C$_2$/CN bands
present) from oxygen-rich stars (TiO bands present) at a glance.

The standard way to measure a band's strength is to compare its flux against
a continuum level interpolated from clean windows well away from any
band. This fails for these spectra: cool carbon and late-M stars are so
densely covered in overlapping molecular bands (i.e., line blanketing) that
there is no clean, band-free window nearby against which to define a continuum; 
windows placed $120$--$150$\,\AA\ from a band, as this method requires, still
sit inside absorption from some other band. Applied to our sample, such a
method recovers only 3/14 known carbon stars and misclassifies 18/48
M-type stars as carbon stars.

\begin{figure*}[t]
\centering
\includegraphics[width=0.92\textwidth]{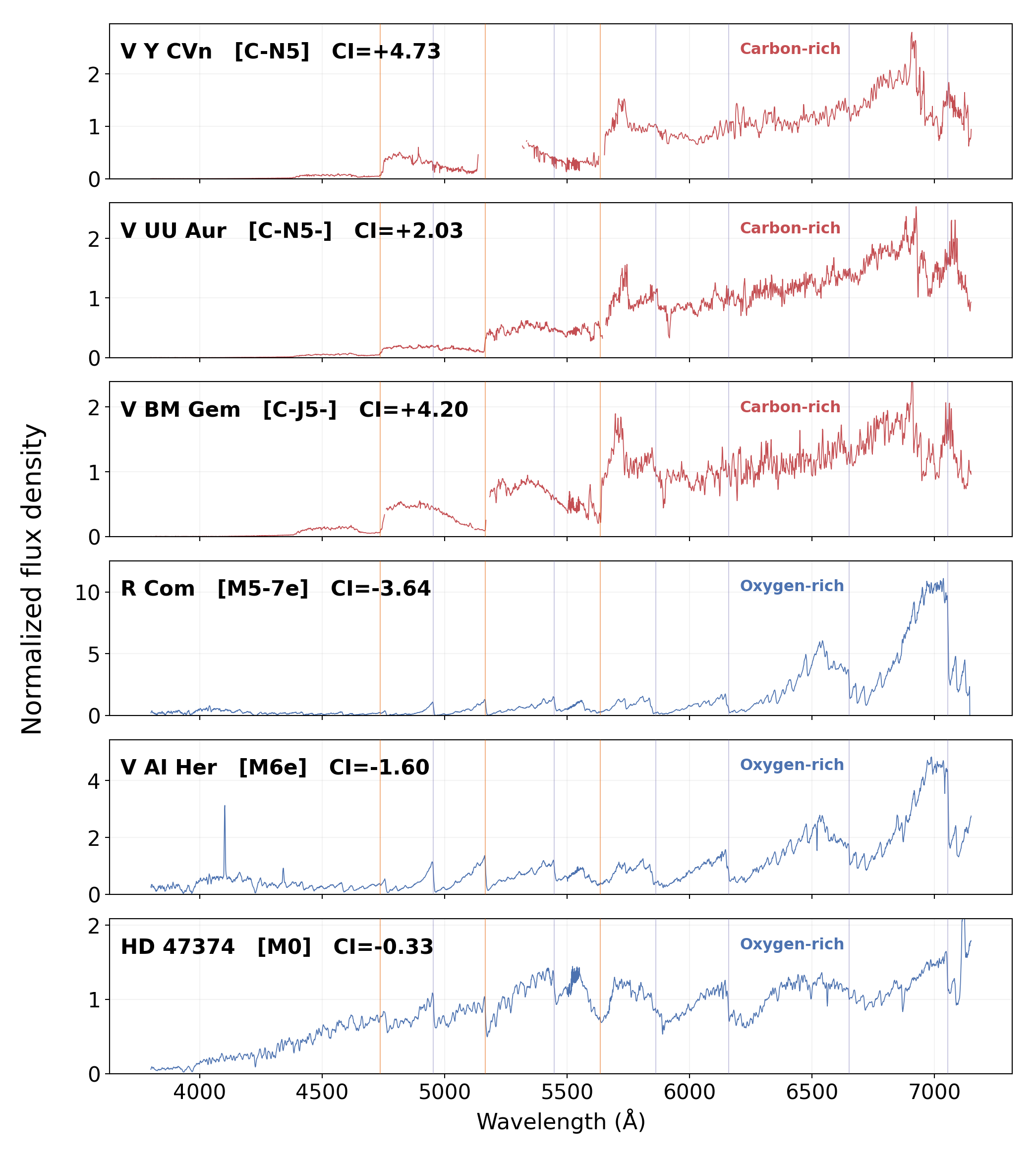}
\caption{Representative Shane/Kast spectra. Carbon-rich (red) panels show the
three highest-S/N carbon stars. Oxygen-rich (blue) panels are chosen for
spectral-subtype diversity (coolest, median, and warmest bandhead-step index
among M-type oxygen-rich stars with $\mathrm{S/N}\ge30$) rather than by S/N
alone, since the highest-S/N oxygen-rich stars are all mid-M giants with
near-identical TiO-band shapes; this instead illustrates how the TiO bands
strengthen with cooler subtype. Vertical lines mark C$_2$ Swan heads (orange)
and TiO heads (purple). The narrow, high-frequency structure visible in the
red arm, especially for the three carbon-rich spectra, is not photon noise:
per-pixel S/N there exceeds 100 throughout, well above what such variations
would require, so this structure reflects real narrow spectral features
and/or residual systematics from the telluric correction
(Section~\ref{sec:reduction}).\label{fig:atlas}}
\end{figure*}

We therefore instead measure \emph{bandhead steps}: rather than referencing a
distant continuum, we directly compare the flux on the two sides of a
bandhead to each other. Bracketing a single feature between flanking
continuum windows has precedent, e.g., the TiO/VO band indices of
\citet{KenyonFernandezCastro1987} for cool giants in symbiotic-star
spectra, but we are not aware of one for the particular application of
simultaneously combining nine such bandhead measurements into a single
classification index. We use six TiO heads
(4954, 5448, 5862, 6159, 6651, 7054\,\AA) and three C$_2$ Swan heads (4737,
5165, 5636\,\AA), with the Swan bands being the characteristic set of C$_2$ optical
bands. For each head we place two narrow windows
immediately on either side, with a
$10$\,\AA\ gap and a $35$\,\AA\ width on each side (i.e.,\ the two windows span
$[-45,-10]$\,\AA\ and $[+10,+45]$\,\AA\ from the head). TiO heads are shaded
(absorption is deeper) on their red side, and C$_2$ Swan heads on their violet
(blue) side (Figure~\ref{fig:atlas} marks these heads on representative
spectra); comparing the flux in the two windows therefore isolates exactly
the asymmetry each molecule produces, without needing a distant continuum
reference at all. Concretely, we form (blue/red)$-1$ for TiO and
(red/blue)$-1$ for C$_2$ at each
head, clip at zero, and average over heads with equal weight rather than by
per-head signal-to-noise. The $0.58$\,mag separation margin achieved below
(Section~\ref{sec:bandhead}) shows this simplification does not compromise
the classification. We classify on a discriminant of
our own naming, the ``chemistry index,''
$\mathrm{CI} = \mathrm{C_2\ index}$ minus $\mathrm{TiO\ index}$ (not to be
confused with a photometric color index): carbon-rich for $\mathrm{CI}>0$,
following the same qualitative M$\to$MS$\to$S$\to$SC$\to$C sequence of
increasing carbon enrichment used in the classification of
\citet{Keenan1954} and \citet{KeenanBoeshaar1980}, who term the analogous
quantity an abundance index rather than a chemistry index. We exclude the
TiO head that would otherwise sit at
5167\,\AA, which is blended with the C$_2$ Swan head at 5165\,\AA\ at Kast
resolution and cannot be attributed to either molecule.

\begin{figure}[t]
\centering
\includegraphics[width=\columnwidth]{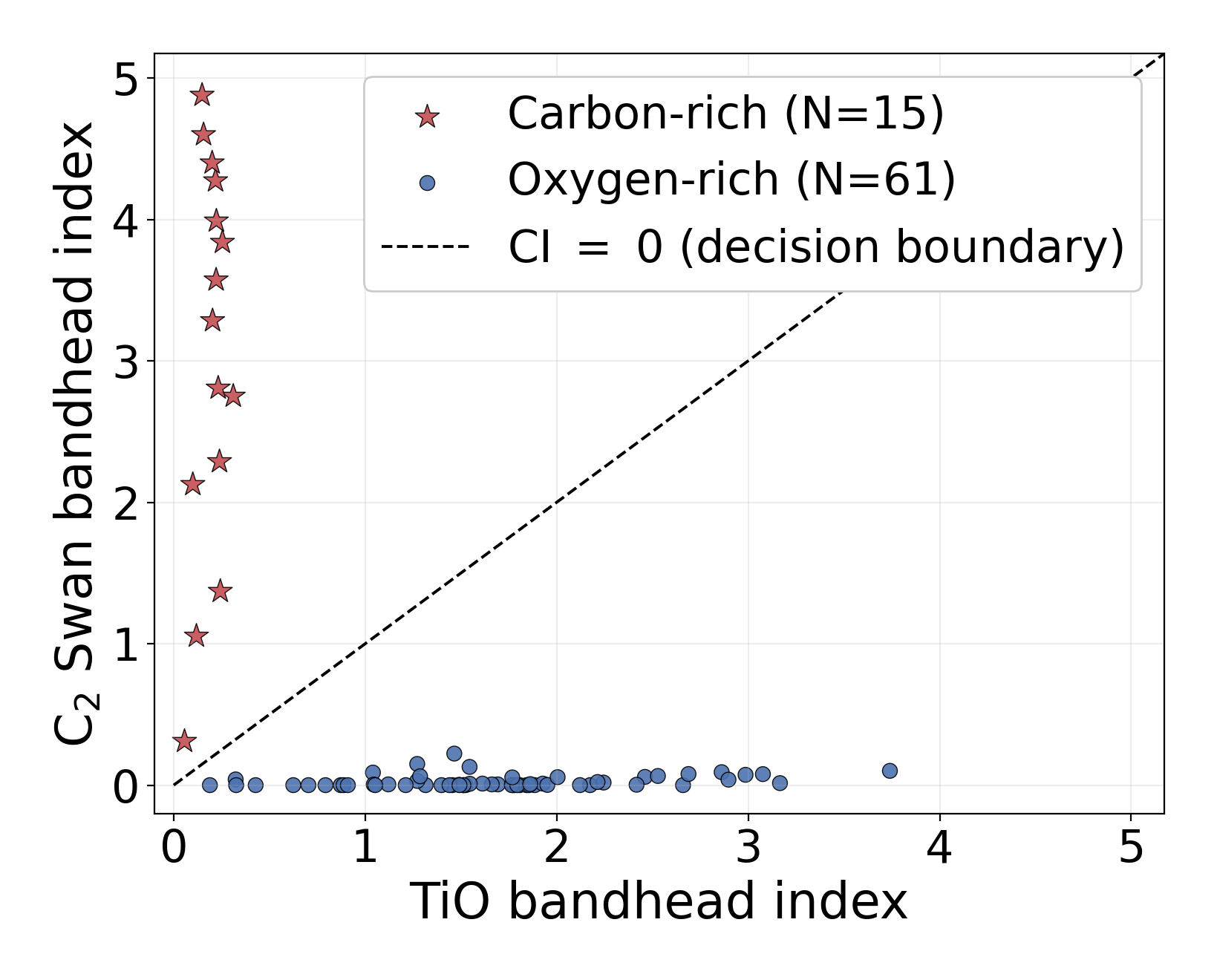}
\caption{Bandhead-step diagnostic plane. Carbon stars occupy the C$_2$ axis and
oxygen-rich stars the TiO axis; the dashed line is the decision boundary
$\mathrm{CI}=0$. The separation is complete for the labeled subset
(14/14 and 48/48).\label{fig:diag}}
\end{figure}

\begin{figure}[t]
\centering
\includegraphics[width=\columnwidth]{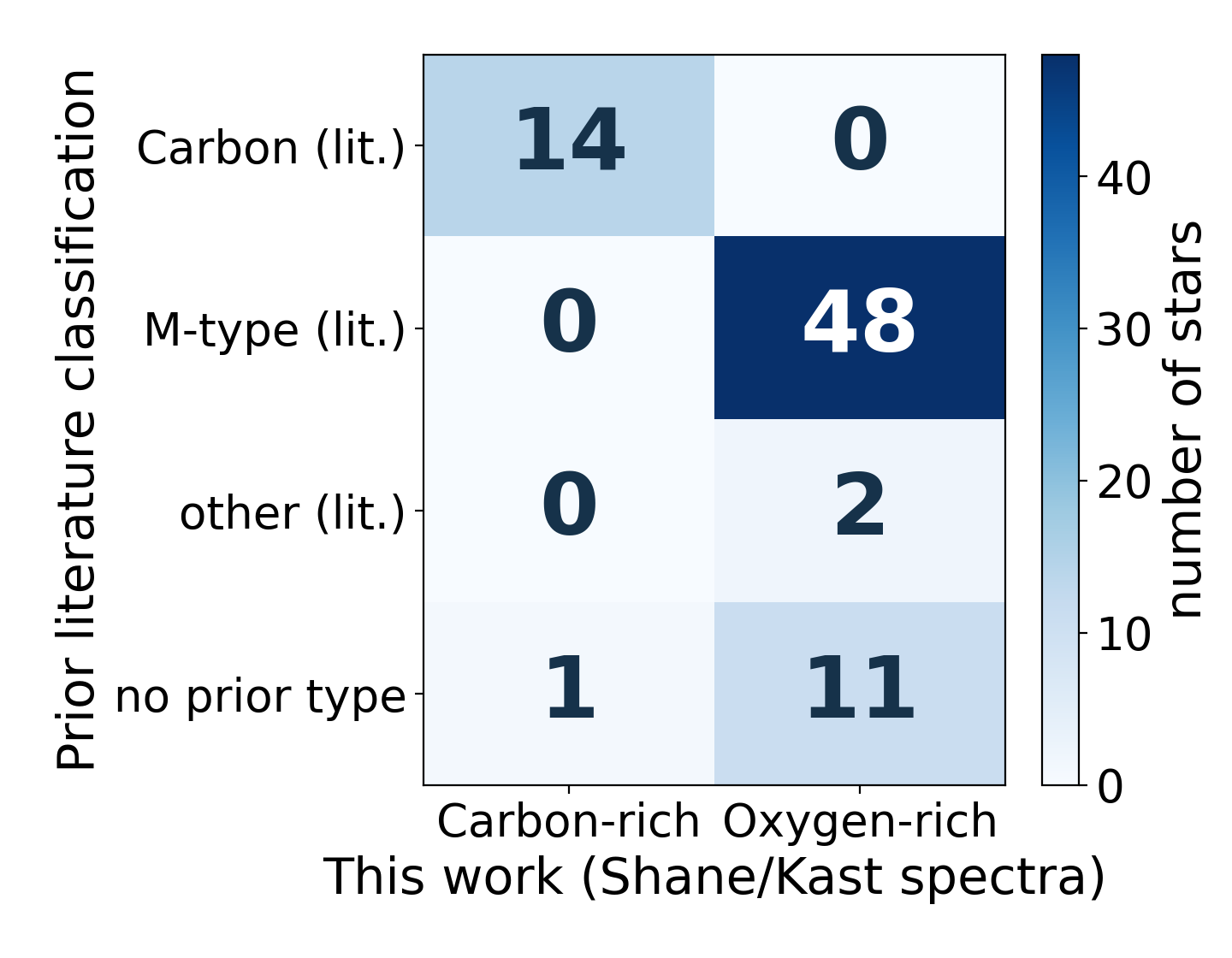}
\caption{Our classification against prior literature types: 14/14 known
carbon and 48/48 known M-type stars recovered. We classify the chemistry of 12 stars without prior classifications in this work. \label{fig:confusion}}
\end{figure}

Against the
SIMBAD-labeled subset (62 of 76 observed stars with a prior SIMBAD carbon or
M-type classification; the remaining two SIMBAD-classified stars, an S-type
and a K5 giant, fall outside both categories and are excluded from this
validation) this recovers 14/14 carbon and 48/48
M-type stars, with
$\min(\mathrm{CI})=+0.253$ for carbon and $\max(\mathrm{CI})=-0.326$ for
M-type, a clean margin (Figs.~\ref{fig:diag} and~\ref{fig:confusion}). We stress
that this validation uses the same stars on which it is demonstrated, so it is not an
independent test. It is, however, not a circular one either: $\mathrm{CI}=0$ is
not a threshold fit to these labels, but the physically motivated zero-crossing
between C$_2$-dominated and TiO-dominated spectra, fixed before comparison to
SIMBAD types; the resulting $0.58$ separation gap between classes emerged from
the data rather than being tuned to produce it. The
ordering is physically sensible: the carbon star nearest the boundary is
RR\,Her, the sample's one SC-type transition object; SC stars are AGB
stars with a carbon-to-oxygen ratio near 1, spectroscopically intermediate
between S-type and carbon stars. Only 1 of the 76 stars in our sample falls
in this transitional class, suggesting that, within this limited sample, SC
stars are not the dominant contaminant in the JAGB region. The M-type stars
nearest the $\mathrm{CI}=0$ boundary are M0, the warmest and weakest in TiO.
Photon-noise-propagated uncertainties, from Monte Carlo resampling of each
pixel's flux within its measured uncertainty, are small relative to this
margin: $\sigma_{\mathrm{CI}}=0.007$ for RR\,Her and $0.002$ for HD\,47374,
the carbon- and M-type stars nearest the boundary, $39\sigma$ and $191\sigma$
from $\mathrm{CI}=0$, respectively, so noise cannot plausibly move either
boundary star across the classification threshold.

As an independent literature cross-check that requires no reprocessing of raw
spectra, we compared our bandhead classifications to two Gaia-based
C-rich definitions: the RP-spectrum classifier of \citet{Lebzelter2023}
(\texttt{isCstar}, set when the median separation between the two highest
peaks of a star's Gaia RP spectrum exceeds a fixed threshold in
Gaia's own pipeline) and the Gaia-2MASS Wesenheit diagram of
\citet{Lebzelter2018} and \citet{Mowlavi2019}, evaluated here using each star's own
absolute magnitude from its Bailer-Jones distance. Of our 76 stars, 57
(75\%) have an entry in the Gaia DR3 long-period-variable catalog; all 57 agree with our classification. The Gaia-2MASS diagram,
computable for all 76 stars independent of long-period-variable status,
agrees for 73/76 (96\%); the three exceptions are listed in
Table~\ref{tab:gaia2m}, none a strong outlier;
ATO\,J106.0289+18.2473 is the one star we newly classify as carbon-rich in
Section~\ref{sec:newclass}, previously only a photometric Gaia RVS
candidate.

\begin{deluxetable}{lcc}
\tabletypesize{\footnotesize}
\tablecaption{Literature spectral-catalog cross-match yield.\label{tab:catalogs}}
\tablehead{\colhead{Catalog} & \colhead{Matched} & \colhead{Retrieved}}
\startdata
Gaia DR3 XP  & 76 & 69 \\
LAMOST DR11           & 19 & 19 \\
Gaia DR3 RVS & -- & 18 \\
Barnbaum et al. (1996)& 5  & 4  \\
IRTF library (NIR)    & 2  & 2  \\
XSL DR3 \citep{Verro2022} & 1 & 1 \\
APOGEE DR17 \citep{Abdurrouf2022} & 4 & 3 \\
MILES, XSL DR1,       & 0  & 0  \\
RAVE, GALAH, SDSS,    &    &    \\
IRAS LRS, PGIR        &    &    \\
\enddata
\tablecomments{\emph{Matched} is the
number of our stars with a positional or identifier cross-match in each
catalog; \emph{Retrieved} is the subset of those matches for which a usable
spectrum was actually downloaded and passed our data-quality checks (S/N,
wavelength coverage). A match can fail to yield a spectrum because of
missing/masked data, non-crossmatch-only catalogs, or a failed query, so
Retrieved is always $\le$ Matched. Gaia DR3 RVS spectra were not
separately cross-matched but drawn from the same Gaia DR3 XP-matched
stars, hence no Matched entry.}
\end{deluxetable}

\begin{table*}[t]
\centering
\caption{Discrepant star classifications.} \label{tab:gaia2m}

\footnotesize
\begin{tabular}{lccc}
\hline
Star & This work & Gaia-2M class & $W_{RP}-W_K$ \\
\hline
ATO J106.0289+18.2473 & Carbon-rich & O-rich (faint AGB/RGB) & $+1.03$ \\
HD 47374 & Oxygen-rich & C-rich (not extreme) & $+1.54$ \\
V T Cnc & Carbon-rich & O-rich (massive AGB/RSG) & $+0.85$ \\
\hline
\end{tabular}
\tablecomments{The three stars where the Gaia-2MASS diagram classification of
\citet{Lebzelter2018} and \citet{Mowlavi2019} disagrees with our bandhead-index
classification. All 57 stars with a Gaia DR3 long-period-variable
catalog entry agree with the \citet{Lebzelter2023} \texttt{isCstar} flag.}
\end{table*}

\subsection{New Classifications}
\label{sec:newclass}
Twelve stars have no spectral type recorded in SIMBAD. We classify them here,
but note that absence from SIMBAD does not strictly prove no earlier
classification exists. We classify eleven as oxygen-rich and
one, ATO\,J106.0289+18.2473, previously only a candidate from the Gaia
RVS (Radial Velocity Spectrometer) carbon-star list, as carbon-rich
($\mathrm{CI}=+3.77$). Our
sample therefore contains a total of 15 carbon stars.

\subsection{Composition of the JAGB Box}
\label{sec:jagbcomp}
We adopt the JAGB color cuts of \citet{Lee2023} and \citet{LeeMagellan2024}:
$1.5<\jk<2.0$ mag. This range also sits inside those adopted by other
J/$(J-K_s)$ JAGB studies, e.g., \citet{MadoreFreedman2020}, \citet{Ripoche2020},
\citet{Lee2021}, \citet{Parada2021}, \citet{Parada2023}, and
\citet{Zgirski2021}, where the red edge remains at $2.0$ mag and the blue
edge can be slightly bluer than $1.5$ mag. We explore the effects of
changing this blue edge later in the paper (Section~\ref{sec:colorcut}).

\begin{figure}[t]
\centering
\includegraphics[width=\columnwidth]{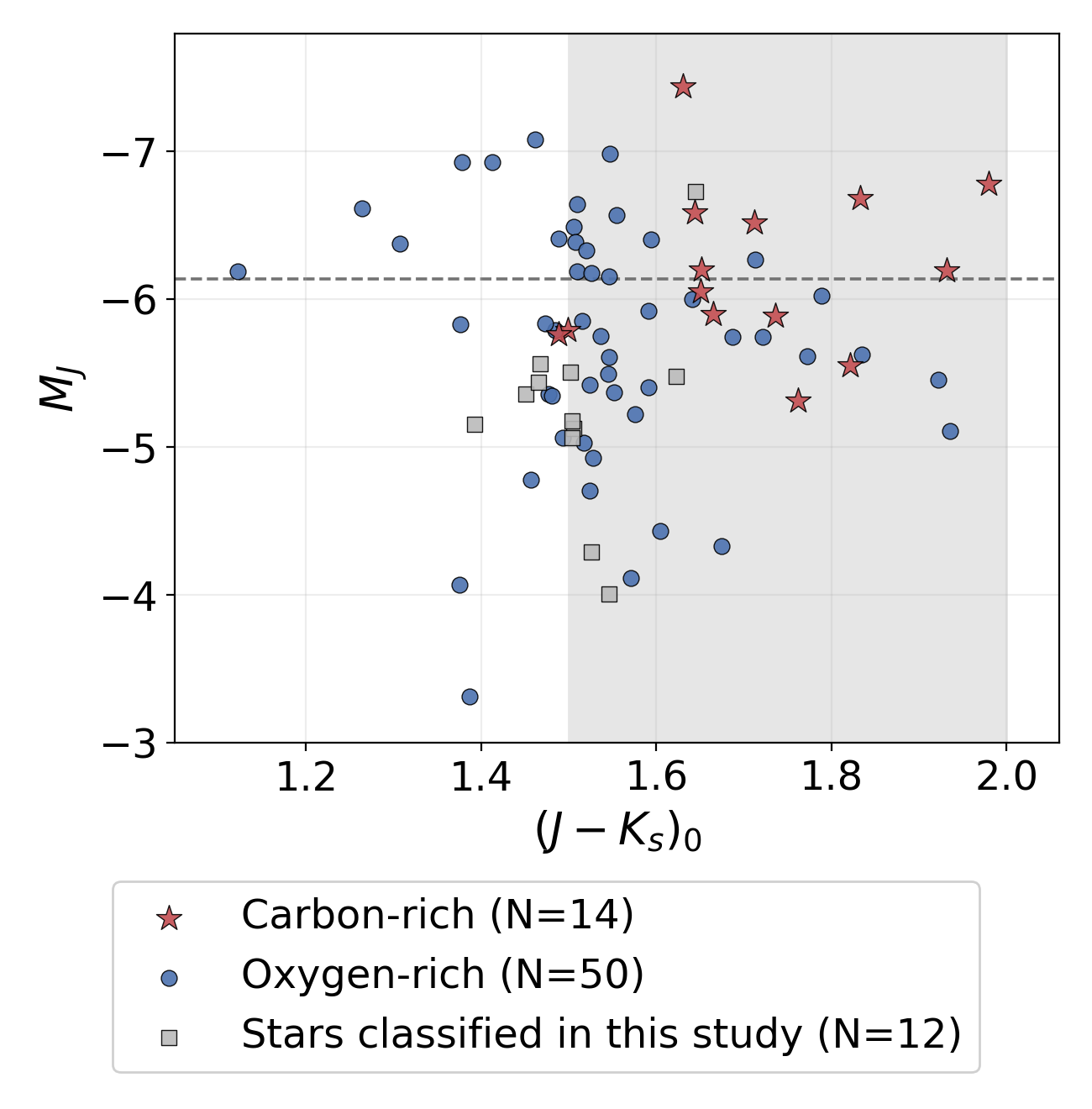}
\caption{Color--magnitude diagram of the observed sample, dereddened with the
Bayestar19 3D dust map \citep{Green2019} and placed on Bailer-Jones distances
\citep{BailerJones2021}. Shading marks the
JAGB color box $1.5<\jk<2.0$ mag; the dashed line is the \citet{Lee2021} zero
point. Chemistry is from prior literature catalogs; stars lacking a prior
spectral type (squares) are those classified spectroscopically in this work.
Most of the box is oxygen-rich, despite the color box being designed to
select carbon-rich stars. Typical photometric uncertainties are $\lesssim0.03$
mag, rising to $\sim0.2$--$0.3$ mag for 2MASS-saturated stars
(Section~\ref{sec:caveats}).\label{fig:cmd}}
\end{figure}

Within $1.5<\jk<2.0$ mag and the parent cuts, 13 of 54 stars are carbon-rich
(2 of the sample's 15 carbon stars lie outside this color box; see
Section~\ref{sec:colorcut}):
purity $24\pm6\%$, corresponding to $76\pm6\%$ oxygen-rich contamination
(Fig.~\ref{fig:cmd}), where the uncertainty is the binomial standard error,
$\sqrt{p(1-p)/N}$. This result is robust:
adopting literature rather than our own classifications gives $76\%$ (zero
disagreements with our Kast classifications among the 46 box stars with a
literature type), and using full-line-of-sight SFD reddening and naive
inverse-parallax distances instead of a 3D map and Bailer-Jones distances
gives $75\%$, all well within the statistical uncertainty.

Restricting to the JAGB peak itself does
not help: within $\pm0.4$\,mag of $\mj=-6.14$ mag the purity is $29\%$ ($6$ of
$21$), and within $\pm0.8$\,mag it is $26\%$ ($10$ of $38$). The
contamination is therefore not confined to the horizontal wings of the selection box.

\subsection{Carbon-star Luminosities}
\begin{figure}[t]
\centering
\includegraphics[width=\columnwidth]{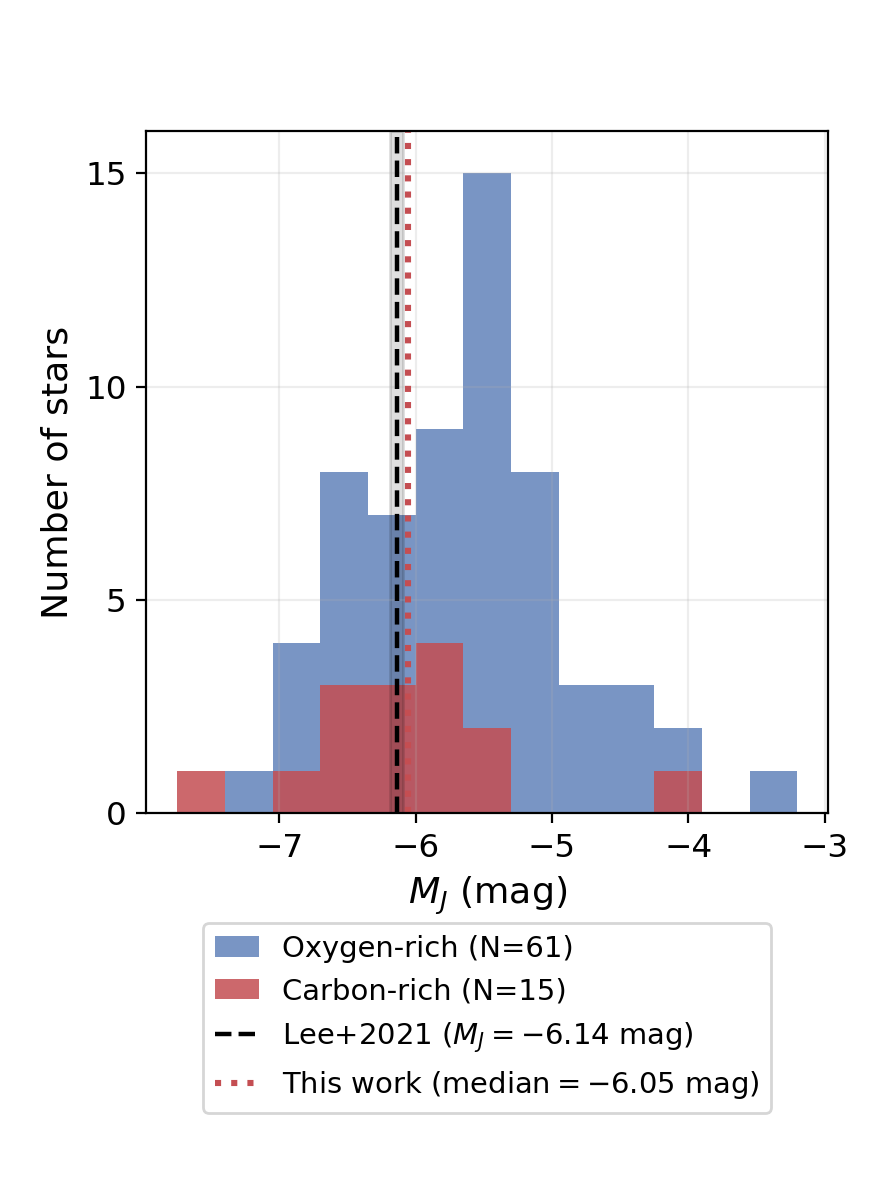}
\caption{$M_J$ distribution split by chemistry. The dashed line and gray band
show the \citet{Lee2021} zero point.\label{fig:lf}}
\end{figure}

For the 15 carbon stars we find $\mj$ = $-6.05$ mag (median), $-6.05\pm0.20$ mag (mean),
$\sigma=0.78$ (Fig.~\ref{fig:lf}), using Bailer-Jones geometric distances and
Bayestar19 reddening. The quoted uncertainty is the standard error on the mean, $\sigma/\sqrt{N}$, and is
statistical only. This value lies between those of \citet{Lee2021} ($-6.14$ mag) and
\citet{Magnus2024} ($-5.90$ mag, weighted mean), but we emphasize that with
$\pm0.20$\,mag, statistical only, it differs from \citet{Lee2021} by only $0.5\sigma$: these data
have no power to discriminate between the published Milky Way values, and we do
not claim to. Single-epoch 2MASS photometry of large-amplitude variables
inflates $\sigma$, and the sample is small.

\begin{figure}[t]
\centering
\includegraphics[width=\columnwidth]{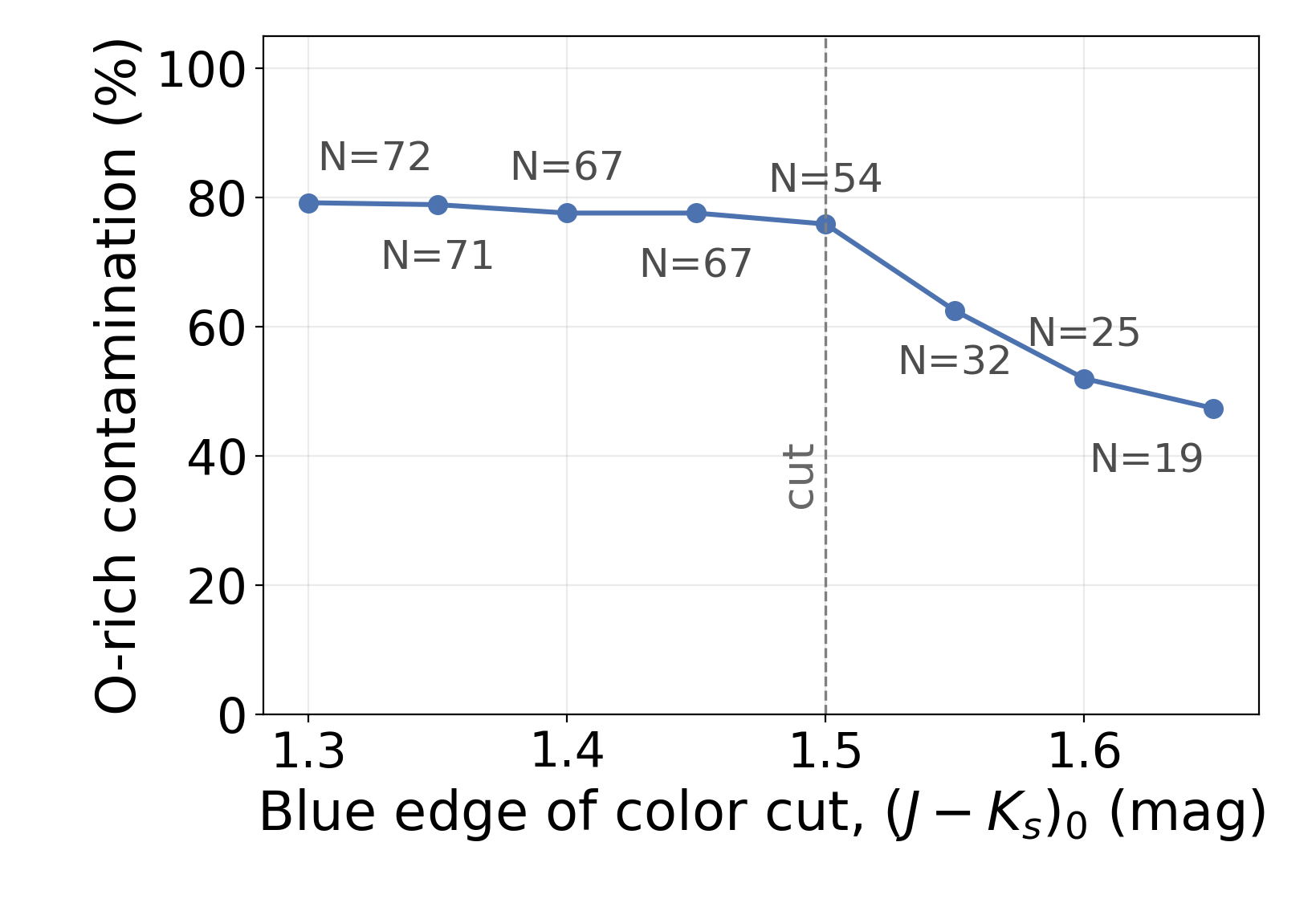}
\caption{O-rich contamination of the box as a function of the blue color cut
(red edge fixed at 2.0 mag); $N$ is annotated at each point.\label{fig:cut}}
\end{figure}

\begin{deluxetable*}{lccc}
\tablecaption{Log of Shane 3\,m/Kast observations.\label{tab:log}}
\tablehead{\colhead{UTC date} & \colhead{UTC start} &
\colhead{$N_{\rm sci}$ (B/R)} & \colhead{$T_{\rm exp}$ (B/R, s)}}
\startdata
2026-02-13 & 05:30 & 49/45 & 2995/2736 \\
2026-03-16 & 02:48 & 171/171 & 6706/6569 \\
2026-05-13 & 03:51 & 98/103 & 8629/5793 \\
\enddata
\tablecomments{All three nights used an identical configuration: 600/4310 grism (blue arm), 600/5000 grating (red arm), d57 dichroic, and a $2.0''$-wide slit. Wavelength coverage measured from the coadded spectra: blue $\sim$3590--5710\,\AA\ at 1.06\,\AA\,pix$^{-1}$, red $\sim$5240--7150\,\AA\ at 1.29\,\AA\,pix$^{-1}$. Exposure times are summed over all science frames.}
\end{deluxetable*}

\begin{deluxetable}{ccccc}
\tablecaption{O-rich contamination of the JAGB color box. \label{tab:contam}} 
\tablehead{\colhead{Blue edge} & \colhead{$N$} & \colhead{$N_{\rm C}$} &
\colhead{Purity (\%)} & \colhead{Contamination (\%)}}
\startdata
1.3 & 72 & 15 & 20.8 & 79.2 \\
1.35 & 71 & 15 & 21.1 & 78.9 \\
1.4 & 67 & 15 & 22.4 & 77.6 \\
1.45 & 67 & 15 & 22.4 & 77.6 \\
1.5 & 54 & 13 & 24.1 & 75.9 \\
1.55 & 32 & 12 & 37.5 & 62.5 \\
1.6 & 25 & 12 & 48.0 & 52.0 \\
1.65 & 19 & 10 & 52.6 & 47.4 \\
\enddata
\tablecomments{Purity and contamination as a function of the blue color cut (red edge fixed at $(J-K_s)_0=2.0$ mag.)}
\end{deluxetable}

\section{Discussion}
\label{sec:discussion}
\subsection{An Inhomogeneous JAGB Region}
The JAGB selection region is not a clean carbon-star sample; three-quarters of the sample examined here is oxygen-rich.
The carbon stars are themselves heterogeneous: the literature types of our 15
carbon stars span N-type, J-type, and SC-type objects, which do not share a
single evolutionary origin. J-type carbon stars in particular are not simple
third-dredge-up products \citep{AbiaIsern2000}. We stress that these subtypes
are taken from published classifications, not measured here: the bandhead index
of Section~\ref{sec:bandhead} separates carbon- from oxygen-rich chemistry but does not by
itself resolve N/J/SC subtypes, and extending it to do so is left to future work.

We suggest, as a hypothesis rather than a result, that a population this mixed
could contribute to the LF asymmetry and field-to-field variation reported by
\citet{Ripoche2020},  \citet{Parada2021}, \citet{Parada2023}, and \citet{Li2024,Li2025}, since the mixture would vary with the star-formation
history and metallicity of each host. We emphasize that we have not demonstrated
this: we show that the \emph{Milky Way} JAGB selection box is mixed, not that the extragalactic
JAGB selection boxes are, and we do not demonstrate that such a mixture reproduces the observed
asymmetry. With our existing sample, oxygen-rich box stars are on average
$0.45$\,mag fainter than carbon-rich box stars ($M_J=-5.63$ vs.\ $-6.09$\,mag;
Mann--Whitney $p=0.04$, Welch $t$-test $p=0.09$), suggesting contamination
could bias a blind box-average luminosity toward fainter values, though the
small carbon-star sample ($N=13$) keeps this only marginally significant. A
larger spectroscopic sample would let us test this more robustly, and
directly resolve what effect any such difference would have on JAGB-based
$H_0$ measurements. Once such a sample exists, we can search for mitigation
strategies. In the meantime, quoted JAGB $H_0$ uncertainties may be
underestimated unless population-diversity-driven scatter is explicitly
built into the error budget or accounted for in a consistent manner, such as done in \citet{Parada2021, Parada2023, Li2024} and \citet{Li2025}.

\subsection{Color-cut Sensitivity}
\label{sec:colorcut}
Contamination falls monotonically as the blue edge is reddened, from $79\%$
at $\jk=1.30$ mag to $52\%$ at $1.60$ mag (Fig.~\ref{fig:cut}; Table~\ref{tab:contam}), but not uniformly:
it is nearly flat blueward of the standard cut ($79\%\to76\%$ over $1.30$--$1.50$ mag) and then
falls sharply just redward of it ($76\%\to52\%$, a $24$-percentage-point drop, over only
$1.50$--$1.60$ mag, as the sample shrinks from $N=54$ to $25$). So the blue limit is a weak
lever only in the bluest part of the box; right at the standard $1.5$ cut, it is a strong one.
Two of our carbon stars lie just blueward of
the standard cut, at $\jk=1.499$ and $1.489$ mag; loosening the edge to $1.45$ mag
recovers both at a cost of about two percentage points in purity, for the sample observed in this study.
We caution that this sample is not a representative draw of stars with
$1.3<\jk<1.5$ mag, since targets were selected in part to probe stars near
the box edges (Section~\ref{sec:targetsel}).

\citet{Magnus2024} report a much larger effect from an analogous box change
in their own Milky Way sample: tightening from their Model 101 box
($1.3<\jk<2.0$ mag, $-7.5<\mj<-5.0$ mag; $47\%$ O-rich) to their Model 102 box
($1.4<\jk<2.0$ mag, $-7.4<\mj<-5.1$ mag; $12\%$ O-rich) drops their contamination
by $35$ percentage points. Applying these same two box definitions to our
own sample gives $78\%$ ($N=64$) and $77\%$ ($N=56$) O-rich
contamination (respectively), a $1.3$ percentage-point change. Reclassifying
the same two boxes with the Gaia-2M-and-LPV2 definition of \citet{Magnus2024}
(Section~\ref{sec:bandhead}), restricted to the subset of our stars their
pipeline could evaluate at all, gives $75\%$ ($N=52$) and $73\%$
($N=45$), respectively.

\subsection{Reconciliation with \citet{Magnus2024}}
\citet{Magnus2024} report Milky Way contamination $<8\%$ for their final,
fully curated model, and $47\%$/$12\%$ for their less-curated Model
101/102 (Section~\ref{sec:colorcut}), in apparent conflict with our $76\%$ for
the same nominal color box. We tested three candidate explanations.
(i) \emph{Astrometric cut}: ours
($\varpi/\sigma_\varpi>10$) is stricter than theirs
($R_{\rm plx}\equiv(\varpi+0.1\,\mathrm{mas})/\sigma_\varpi\ge5$, their
zero-point-corrected parallax significance), but the
four stars we reject all lie blueward of the box, so relaxing it does not
change the in-box contamination at all, and thus rejected. (ii) \emph{Color cut}: the trend is in the
right direction but far too weak, and by $\jk=1.60$ mag our sample is only $N=25$
(Section~\ref{sec:colorcut} quantifies this directly against
the Model 101/102 box change of \citet{Magnus2024}), and thus rejected.
(iii) \emph{Classification methodology}: reclassifying our JAGB-box stars
with the Gaia-2M-and-LPV2 C-rich definition of \citet{Magnus2024} instead of
our bandhead index, restricted to the 41 of 54 box stars with an entry in
the Gaia DR3 long-period-variable catalog at all
(Section~\ref{sec:data}), gives $73\%$ O-rich contamination ($30$ of $41$
stars) and is consistent with our own $76\%$. Framed this way, our finding and
that of \citet{Magnus2024} are complementary rather than conflicting: both
studies conclude the Milky Way JAGB selection is not chemically pure, and
neither confirms the metallicity independence the method assumes.

The difference is in what is being measured. Their Milky Way sample is drawn from
\texttt{gaiadr3.vari\_summary}, the Gaia long-period-variable
catalog, so only variable stars enter; chemistry is then assigned from the
Gaia-2M diagram or the LPV2 classification, and their recommended model further
restricts to curated open-cluster, wide-binary, and SAAO-photometry stars. Their
value is therefore the contamination \emph{remaining after} a prior
photometric classification of a variability-selected sample. Our
result, by contrast, measures the contamination in a blind, magnitude-limited
field selection with no prior chemical information. This is the mode of
selection, though not the stellar population, that extragalactic JAGB work
must use, since spectroscopic vetting is impossible at those distances.

The 54 JAGB-box stars are not all sources that the Milky Way selection pipeline of \citet{Magnus2024}, which draws its sample from the Gaia
DR3 long-period-variable catalog (Section~\ref{sec:discussion}), could
have identified in the first place. Only 41 of the 54 (76\%) have any entry
at all in \texttt{gaiadr3.vari\_long\_period\_variable}. The remaining 13
stars are therefore absent from an \citet{Magnus2024}-style sample by
construction, independent of their true chemistry. All 54 pass the astrometric cut of
\citet{Magnus2024} ($R_{\rm plx}\ge5$; see
Section~\ref{sec:discussion} for its definition), which is looser than our
own $\varpi/\sigma_\varpi>10$.

This incompleteness compounds with a further selection effect that
\citet{Magnus2024} note themselves: their photometric input requires a
2MASS quality flag of ``AAA'' in $J$, $H$, and $K_s$, which they caution
likely excludes saturated, nearby AGB stars. Cross-matching our 54 JAGB-box
stars against the 2MASS Point Source Catalog, only 2 carry an AAA quality
flag in all three bands; the remaining 52 would fail this cut and so could
not enter a \citet{Magnus2024}-style sample regardless of their true
chemistry. Because our targets
were selected to be bright enough for ground-based optical spectroscopy,
this AAA requirement removes nearly all of them, offering a further
candidate explanation for why \citet{Magnus2024}'s reported Milky Way
contamination fraction differs from the fraction found here.

\subsection{How Far Does This Generalize?}
\label{sec:scope}
Our measurement applies specifically to the solar neighborhood, and should
not be read as a universal JAGB contamination fraction. The C to late-M
ratio is strongly metallicity dependent, rising from $\sim0.2$ near solar to
$\sim1$ in the LMC and $\sim5$ in the SMC \citep[][and references
therein]{Magnus2024}. Converting this general population ratio into an
expected contamination fraction for a specific photometric selection box
would require assumptions, untested here, about how consistently that box
samples the carbon and late-M populations across environments of differing
metallicity; we therefore do not extrapolate our measurement to other
metallicities.

However, what does transfer is the mechanism rather than the number: a purely photometric
color--magnitude selection admits oxygen-rich AGB stars at a rate set by the
host's C/M ratio (the carbon-to-late-M-type star ratio defined above), and that rate is not small wherever the metallicity approaches
solar. Since the JAGB method is increasingly applied to metal-rich spiral hosts,
quantifying this contamination as a function of host metallicity is, in our
view, a necessary next step, and one that requires spectroscopy of the kind
presented here in environments beyond the solar neighborhood.

\subsection{A Possible Photospheric Contribution to the O-rich Contamination}
\label{sec:comarcs}
\begin{figure}[t]
\centering
\includegraphics[width=\columnwidth]{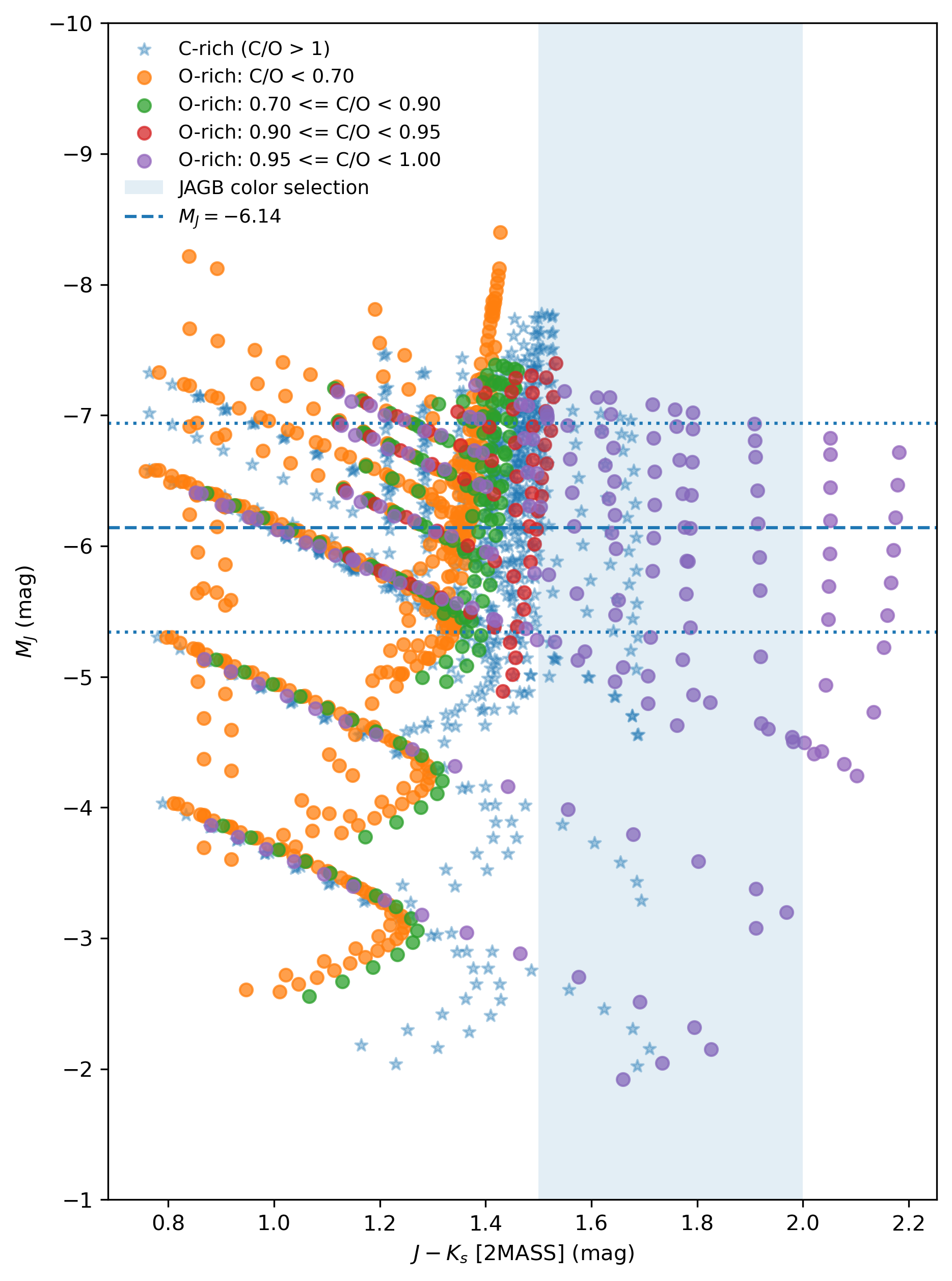}
\caption{COMARCS/COMA hydrostatic, dust-free synthetic model atmospheres
\citep{Aringer2016, Aringer2019} in the $\mj$--$\jk$ plane, split strictly by
carbon-to-oxygen ratio (star = carbon-rich, C/O $>1$; circles = oxygen-rich,
C/O $<1$, subdivided into four C/O bins). The shaded region is the JAGB color
box and the dashed line the \citet{Lee2021} zero point; the dotted horizontal lines indicate $\pm0.8$ mag around this zero point. Both chemistries
populate the JAGB box at the photospheric level, and the oxygen-rich models
that do so cluster near the C/O $=1$ transition. Model counts in this grid are
not population-weighted and should not be read as a predicted contamination
fraction (Section~\ref{sec:comarcs}).\label{fig:comarcs}}
\end{figure}

The contamination reported above is seen to be a photometric selection effect. To test whether the chemical overlap identified above can arise at the
photospheric level, we compare the observed JAGB selection with synthetic
photometry from the COMARCS/COMA atmosphere grid
\citep{Aringer2016,Aringer2019}. COMARCS provides spherical, hydrostatic
model atmospheres calculated assuming local thermodynamic and chemical equilibrium, while
COMA spectral synthesis provides bolometric corrections for the 2MASS
photometric system. The models considered here are dust-free and therefore
do not include reddening or emission from circumstellar dust. We adopt the BT2 H$_2$O molecular line-list treatment \citep{Barber2006} where available, consistent with the synthetic photometry generally presented by \citet{Aringer2016} for cool O-rich models, and otherwise retain the standard COMARCS calculation. Repeating the analysis using only the standard COMARCS treatment does not change our qualitative results presented below. We classify the models by splitting the
grid strictly by carbon-to-oxygen ratio into oxygen-rich (C/O $<1$) and
carbon-rich (C/O $>1$) models and placing both in the $\mj$--$\jk$ plane, as seen in Fig.~\ref{fig:comarcs}.

Both chemical classes enter the conventional $1.5~<~J-K_s~<~2.0$ mag JAGB color
interval in the hydrostatic models, including at luminosities around the
\citet{Lee2021} JAGB zero point. Thus, the JAGB color selection is not
unique to C-rich photospheres even before circumstellar dust is included.
The O-rich overlap, however, occupies a restricted region of the sampled
atmospheric parameter space. We identify that all of the 89 O-rich 
grid points within $1.5<J-K_s<2.0$ mag have C/O $\geq0.90$:
80 have $0.95\leq{\rm C/O}<1$, while the remaining nine have
$0.90\leq{\rm C/O}<0.95$. No sampled O-rich model with C/O $<0.90$
enters the JAGB color interval. The overlapping O-rich models are also
confined to very low effective temperatures
($T_{\rm eff}=\,$2500--3000 K). This behavior is
consistent with the strong changes in molecular equilibrium and
near-infrared colors that occur in cool atmospheres as C/O approaches
$1$ \citep{Aringer2016}.


The models therefore demonstrate a possible photospheric route by which
an atmosphere that is formally O-rich can satisfy the JAGB
color--magnitude criteria without invoking circumstellar dust. At the
same time, the restricted atmospheric regime of the overlap limits how
directly this result can be associated with the contaminants identified
spectroscopically in our sample. In particular, the COMARCS overlap is
concentrated very close to C/O $=1$, whereas our spectroscopic classification
distinguishes O-rich and C-rich stars primarily through TiO and C$_2$
absorption and does not provide a quantitative measurement of C/O. As C/O approaches $1$, the atmospheric chemistry approaches the transition between
M-, S-, and carbon-star regimes, where changes in s-process abundances may also affect the spectrum
\citep{Aringer2016}. Such abundance patterns are not fully represented by
varying C/O alone in the models considered here. Determining whether the
observed contaminants occupy this C/O~$\simeq 1$ regime would therefore
require dedicated S/SC spectral diagnostics or independent abundance
constraints beyond the chemistry index of Section~\ref{sec:bandhead}; testing
this directly against our existing Kast spectra is a natural extension of
this work.

Finally, the relative numbers of O-rich and C-rich models in
Fig.~\ref{fig:comarcs} should not be interpreted as a predicted
contamination fraction. COMARCS is an atmosphere grid rather than a
population-synthesis calculation, and its sampling in C/O, effective
temperature, surface gravity, and other parameters is not weighted by
stellar evolutionary lifetimes or population frequencies. A quantitative
prediction of the O-rich fraction would therefore require coupling the
atmosphere calculations to an appropriate stellar-population model. As a
hydrostatic grid, COMARCS captures only the physics included in its
equilibrium calculation and does not model dynamical processes such as
pulsation-driven shocks or mass loss.

\subsection{Caveats}
\label{sec:caveats}
Single-epoch 2MASS photometry of variable AGB stars inflates the $\mj$ scatter,
and we can bound the effect: for a sinusoidal light curve of full amplitude
$\Delta J$ sampled at random phase, the contributed dispersion is
$\Delta J/2\sqrt{2}$, which is respectively $0.18$, $0.35$, and $0.53$\,mag for
$\Delta J = 0.5$, $1.0$, and $1.5$\,mag,  bracketing the $\sim0.7$\,mag
average $J$-band amplitude adopted by \citet{MadoreFreedman2020}. Even the most extreme of these,
subtracted in quadrature from the observed $\sigma=0.78$\,mag, leaves
$\gtrsim0.57$\,mag of intrinsic and distance-driven scatter, so variability
alone cannot account for the width of our distribution. The same effect applies
to extragalactic single-epoch JAGB work. This scatter could in principle also
move stars across the $\mj$ box edges, not just widen the luminosity
distribution: no star currently in the box sits within $0.53$\,mag (the most
extreme bound above) of either $M_J$ boundary, the closest is $0.87$\,mag
away, so single-epoch variability cannot move stars into or out of the box
for our sample.

The $\mj$ distribution is shaped by the absolute-magnitude selection window,
which is why we do not present it as a calibration. The window spans 6\,mag,
symmetric about the \citet{Lee2021} peak ($-9.14<\mj<-3.14$ mag), whereas the
stars occupy $-7.4$ to $-4.0$ mag, so the
distribution is not truncated by the selection; nevertheless, the window was
applied using preliminary photometry, and a fully unbiased zero point would
require a magnitude-complete sample. With $N=15$ carbon stars, subtype-dependent
luminosity trends are suggestive only. Formal Gaia parallax
uncertainties are optimistic for large-amplitude AGB variables, which our
quoted statistical uncertainty does not capture.

2MASS photometry for many of the sample's brightest, nearest stars is
saturation-limited: 8 of the 15 carbon stars (all within $\sim1.3$\,kpc) and
24 of 26 SIMBAD-confirmed Miras (out to $\sim1.8$\,kpc) carry a 2MASS read
flag of `3' in $J$, $H$, and $K_s$: the point-spread function core is
saturated even in 2MASS's shortest (51\,ms) exposure, so flux is instead
estimated by fitting a profile to the unsaturated scattered-light wings of
the stellar image and extrapolating that fit to recover the total,
core-included flux (``Read\_3'' photometry), a technique 2MASS's
documentation reports successfully photometering sources as bright as
magnitude $-4$\footnote{2MASS All-Sky Data Release Explanatory Supplement,
Section 6.3a.4, \url{https://irsa.ipac.caltech.edu/data/2MASS/docs/releases/allsky/doc/sec6_3a4.html}:
``reliable fluxes could be extracted for sources, even as bright as
magnitude $-4$ (e.g., $\alpha$ Ori and $\alpha$ Tau), using the scattered
light wings in the images... The biases, expected to be $<20\%$ in the
worst cases, lie within the quoted large uncertainties for these
sources.''}. This flux regime falls
outside 2MASS's formal accuracy requirements, with typical uncertainties of
$\sim0.2$--$0.3$\,mag rather than the $\lesssim0.03$\,mag typical of
unsaturated sources. To test whether this affects the JAGB box contamination
fraction above, we perturbed each of the 72 parent-sample stars' dereddened
$\jk$ and $\mj$, not just the 54 in-box stars, so bluer interlopers could
scatter in, each using its own formal 2MASS uncertainty, (20{,}000
realizations) and recomputed box membership and purity each time: the mean
contamination is $73\pm5\%$, statistically consistent with the reported
$76\pm6\%$, with no evidence of a systematic offset from photometric noise
alone. This test is necessarily limited to our 72 observed parent-sample
stars, which are themselves color-selected ($\mathrm{BP-RP}\ge1.3$;
Section~\ref{sec:data}) rather than a blind draw of the much larger field
oxygen-rich population. This test therefore cannot capture oxygen-rich
interlopers scattering in from that excluded, bluer population; since
carbon stars are intrinsically red and rarely fall outside this color cut
to begin with, adding that channel could only increase the number of
simulated oxygen-rich box crossings, not carbon-rich ones, so our
noise-only contamination estimate is, if anything, a lower bound. A natural
follow-up is to repeat it against the full parent-selection catalog of
Section~\ref{sec:data}, and to extend it to
fainter, JAGB-box-selected targets whose 2MASS photometry is more robust,
removing this systematic entirely.

As a further check, we modeled the documented Read\_3 seeing bias directly: for
each of the 32 carbon and Mira stars carrying rd\_flg$=3$ in all three bands,
we perturbed $J$ and $K_s$ independently by up to $0.24$\,mag (the
flux-equivalent of 2MASS's quoted $<20\%$ worst-case bias), drawing a shared
sign per realization so the perturbation acts in the same direction in both
bands while only partially canceling in $\jk$ (20{,}000 realizations). At the
fiducial $\jk>1.5$ cut used throughout this paper ($N=54$), the mean
contamination is $76.4\pm1.4\%$, consistent with the reported $75.9\%$ and
with the noise-only test above: our headline result is robust to this
systematic. The more aggressive cuts explored in Section~\ref{sec:colorcut}
are more sensitive to this bias, with contamination shifting by several
percentage points and substantially larger scatter (e.g., $59.5\pm4.1\%$ at
$\jk>1.6$ versus a baseline of $52.0\%$), but those cuts are well away from
the fiducial value used for our main results.

We test how representative the 54 observed box stars are of the full
parent-sample box population (165 stars meeting the same Gaia/2MASS
selection, Section~\ref{sec:data}) using two-sample KS tests. The observed
sample is statistically indistinguishable from the parent population in both
$\jk$ color ($D=0.09$, $p=0.89$) and $\mj$ ($D=0.07$, $p=0.98$), the two axes
that define the box and enter the purity measurement directly. It is not
representative in sky position: observed stars skew to higher Galactic
latitude (median $b=32^\circ$ vs.\ $19^\circ$ for the parent population,
$D=0.29$, $p=0.001$ by KS test), a consequence of telescope-time and
scheduling constraints (Section~\ref{sec:data}) rather than any astrophysical
cut.

The parent sample is not sky-complete: reddening is drawn from the Bayestar19
3D dust map \citep{Green2019}, built from Pan-STARRS1 photometry and covering
only $\mathrm{Dec}\gtrsim-30^\circ$. Stars south of this limit have no
reddening estimate and are dropped before the dereddened box definition is
ever applied, regardless of whether they would otherwise qualify.
Cross-matching the independent General Catalogue of Galactic Carbon Stars
\citep[CGCS;][]{Alksnis2001} against our exact parent-sample cuts
($b\ge10^\circ$, $G\le13$ mag, $\mathrm{BP}-\mathrm{RP}\ge1.3$ mag,
$\varpi/\sigma_\varpi>10$) finds 152 CGCS stars that should qualify; only 118
(78\%) appear in the parent sample, and 33 of the 34 missing stars lie at
$\mathrm{Dec}<-30^\circ$: a boundary of the Bayestar19 footprint, not a
correctable step within the current pipeline. The remaining star lacks a
Gaia-to-2MASS crossmatch entirely and is dropped before the reddening step
for an unrelated reason (crossmatched at a 5\arcsec\
radius).

\section{Conclusions}
\label{sec:conclusions}
\begin{enumerate}
\item A blind, photometry-only JAGB color-box selection in the solar
      neighborhood is $76\pm6\%$ contaminated by oxygen-rich stars, robust
      to the reddening treatment and classification source.
      The contamination remains high ($71\%$) near the JAGB peak itself.
\item The corresponding carbon fraction is $24\pm6\%$, measured specifically
      for this solar-neighborhood sample; Section~\ref{sec:scope} discusses
      why we do not extrapolate this value to other metallicity
      environments.
\item While band-depth classification fails on blanketed cool-star spectra; bandhead-step indices recover 14/14 carbon and 48/48 M-type stars.
\item We provide chemistry classifications for 12 Milky Way stars lacking a SIMBAD
      spectral type, including one newly confirmed carbon star.
\item CHASE is an ongoing survey; we plan to collect additional
      spectroscopic observations to improve the statistical power of these
      conclusions.
\end{enumerate}

\begin{acknowledgments}
S.L. acknowledges support for a postdoctoral fellowship from the Brinson Foundation.
A.V.F.’s research group at U.C. Berkeley is supported by Timothy and Melissa Draper, Briggs and Kathleen Wood, Ellyn and Alan Seelenfreund (T.G.B. is Draper-Wood-Seelenfreund Specialist in Astronomy), and 
numerous other donors.
A major upgrade of the Kast spectrograph on the Shane 3\,m telescope at Lick Observatory, led by Brad Holden, was made possible through gifts from the Heising-Simons Foundation, William and Marina Kast, and the University of California Observatories. We appreciate the expert  assistance of the staff at Lick Observatory. Research at Lick Observatory is partially supported by a gift from Google. AA and NM acknowledge the support from NAOJ JASMINE Scientific Research Grant Code 2024-01. S.L. thanks Natalie LeBaron for helpful guidance on the use of PypeIt.

Claude (Anthropic; Opus 5 and
Sonnet 5) \citep{Anthropic2026} was used to assist with this paper.
This work makes use of \texttt{PypeIt} \citep{PypeIt}, \texttt{astropy},
\texttt{astroquery}, and \texttt{dustmaps}; the SIMBAD database and VizieR
service (CDS, Strasbourg); ESA Gaia and DPAC; 2MASS \citep{TwoMASS};
and LAMOST.

\end{acknowledgments}

\facilities{Shane (Kast), Gaia, CTIO:2MASS, LAMOST}
\software{PypeIt \citep{PypeIt}, astropy, astroquery, dustmaps, numpy,
          matplotlib, scipy, pandas}

\bibliographystyle{aasjournalv7.1}
\bibliography{refs}

@ARTICLE{Planck2020,
  author  = {{Planck Collaboration}},
  title   = {{Planck 2018 results. VI. Cosmological parameters}},
  journal = {\aap}, year = 2020, volume = {641}, pages = {A6},
  eprint  = {1807.06209}
}

@ARTICLE{Casertano2026,
  author  = {{Casertano}, Stefano and {Anand}, Gagandeep and
             {Anderson}, Richard I. and {Beaton}, Rachael and
             {Bhardwaj}, Anupam and {Blakeslee}, John P. and
             {Boubel}, Paula and {Breuval}, Louise and {Brout}, Dillon and
             {Cantiello}, Michele and {Cruz Reyes}, Mauricio and
             {Cs{\"o}rnyei}, Geza and {de Jaeger}, Thomas and
             {Di Valentino}, Eleonora and {Galbany}, Llu{\'\i}s and
             {Gil-Mar{\'\i}n}, H{\'e}ctor and {Graczyk}, Dariusz and
             {Huang}, Caroline and {Jensen}, Joseph B. and
             {Kervella}, Pierre and {Leibundgut}, Bruno and
             {Lengen}, Bastian and {Li}, Siyang and {Macri}, Lucas and
             {{\"O}z{\"u}lker}, Emre and {Pesce}, Dominic W. and
             {Riess}, Adam and {Romaniello}, Martino and {Said}, Khaled and
             {Sch{\"o}neberg}, Nils and {Scolnic}, Dan and
             {Sicignano}, Teresa and {Skowron}, Dorota M. and
             {Uddin}, Syed A. and {Verde}, Licia and {Nota}, Antonella},
  title   = {{The Local Distance Network: A community consensus report on
             the measurement of the Hubble constant at $\sim$1\% precision}},
  journal = {\aap}, year = 2026, volume = {708}, pages = {A166},
  doi     = {10.1051/0004-6361/202557993}, eprint = {2510.23823}
}

@ARTICLE{NikolaevWeinberg2000,
  author  = {{Nikolaev}, Sergei and {Weinberg}, Martin D.},
  title   = {{Stellar Populations in the Large Magellanic Cloud from 2MASS}},
  journal = {\apj}, year = 2000, volume = {542}, pages = {804},
  doi     = {10.1086/317048}
}

@ARTICLE{WeinbergNikolaev2001,
  author  = {{Weinberg}, Martin D. and {Nikolaev}, Sergei},
  title   = {{Structure of the Large Magellanic Cloud from 2MASS}},
  journal = {\apj}, year = 2001, volume = {548}, pages = {712},
  doi     = {10.1086/319001}
}

@ARTICLE{Alksnis2001,
  author  = {{Alksnis}, A. and {Balklavs}, A. and {Dzervitis}, U. and
             {Eglitis}, I. and {Paupers}, O. and {Pundure}, I.},
  title   = {{General Catalog of Galactic Carbon Stars by C.~B. Stephenson.
             Third Edition}},
  journal = {Baltic Astronomy}, year = 2001, volume = {10}, pages = {1},
  doi     = {10.1515/astro-2001-1-202}
}

@ARTICLE{Gonneau2016,
  author  = {{Gonneau}, A. and {Lan{\c{c}}on}, A. and {Trager}, S.~C. and
             {Aringer}, B. and {Lyubenova}, M. and {Chen}, Y.-P. and
             {Peletier}, R.~F. and {Prugniel}, Ph. and {Silva}, D.~R.},
  title   = {{Carbon stars in the X-Shooter Spectral Library}},
  journal = {\aap}, year = 2016, volume = {589}, pages = {A36},
  doi     = {10.1051/0004-6361/201526292}, eprint = {1602.00887}
}

@ARTICLE{Verro2022,
  author  = {{Verro}, K. and {Trager}, S.~C. and {Peletier}, R.~F. and
             {Lan{\c{c}}on}, A. and {Gonneau}, A. and {Vazdekis}, A. and
             {Prugniel}, Ph. and {Chen}, Y.-P. and {Coelho}, P.~R.~T. and
             {Sanchez-Blazquez}, P. and {Martins}, L. and {Arentsen}, A. and
             {Lyubenova}, M. and {Falcon-Barroso}, J. and {Dries}, M.},
  title   = {{The X-shooter Spectral Library (XSL): Data Release 3}},
  journal = {\aap}, year = 2022, volume = {660}, pages = {A34},
  doi     = {10.1051/0004-6361/202142388}, eprint = {2110.10188}
}

@ARTICLE{Sloan2025,
  author  = {{Sloan}, G.~C. and {Kraemer}, Kathleen E. and {Volk}, K.},
  title   = {{The Extended Atlas of Low-resolution Spectra from the Infrared
             Astronomical Satellite}},
  journal = {\apjs}, year = 2025, volume = {279}, pages = {15},
  doi     = {10.3847/1538-4365/add7d6}, eprint = {2507.14134}
}

@ARTICLE{Abdurrouf2022,
  author  = {{Abdurro'uf} and {Accetta}, K. and {Aerts}, C. and others},
  title   = {{The Seventeenth Data Release of the Sloan Digital Sky Surveys:
             Complete Release of MaNGA, MaStar, and APOGEE-2 Data}},
  journal = {\apjs}, year = 2022, volume = {259}, pages = {35},
  doi     = {10.3847/1538-4365/ac4414}, eprint = {2112.02026}
}

@ARTICLE{Earley2025,
  author  = {{Earley}, Nicholas and {Karambelkar}, Viraj and {Kasliwal}, Mansi and
             {De}, Kishalay and {Hillenbrand}, Lynne and {Soria}, Roberto and
             {Suresh}, Aswin and {Ashley}, Michael C.~B. and {Hankins}, Matthew J. and
             {Moore}, Anna M. and {Soon}, Jamie and {Travouillon}, Tony},
  title   = {{A Spectral Library and Census of Near-infrared Stellar
             Large-amplitude Variables from Palomar Gattini-IR}},
  journal = {\pasp}, year = 2025, volume = {137}, pages = {114205},
  doi     = {10.1088/1538-3873/ae170d}, eprint = {2510.18959}
}

@ARTICLE{MadoreFreedman2020,
  author  = {{Madore}, B.~F. and {Freedman}, W.~L.},
  title   = {{Astrophysical Distance Scale: The AGB J-band Method. I.
             Calibration and a First Application}},
  journal = {\apj}, year = 2020, volume = {899}, pages = {66},
  doi     = {10.3847/1538-4357/aba045}, eprint = {2005.10792}
}

@ARTICLE{FreedmanMadore2020,
  author  = {{Freedman}, W.~L. and {Madore}, B.~F.},
  title   = {{Astrophysical Distance Scale. II. Application of the JAGB Method:
             A Nearby Galaxy Sample}},
  journal = {\apj}, year = 2020, volume = {899}, pages = {67},
  doi     = {10.3847/1538-4357/aba9d8}, eprint = {2005.10793}
}

@ARTICLE{Lee2021,
  author  = {{Lee}, A.~J. and {Freedman}, W.~L. and {Madore}, B.~F. and
             {Owens}, K.~A. and {Sung Jang}, I.},
  title   = {{A Preliminary Calibration of the JAGB Method Using Gaia EDR3}},
  journal = {\apj}, year = 2021, volume = {923}, pages = {157},
  doi     = {10.3847/1538-4357/ac2f4c}, eprint = {2110.04576}
}

@ARTICLE{Lee2023,
  author  = {{Lee}, A.~J.},
  title   = {{Carbon Stars as Standard Candles: An Empirical Test for the
             Reddening, Metallicity, and Age Sensitivity of the J-region
             Asymptotic Giant Branch (JAGB) Method}},
  journal = {\apj}, year = 2023, volume = {956}, pages = {15},
  doi     = {10.3847/1538-4357/acee69}, eprint = {2305.02453}
}

@ARTICLE{Ripoche2020,
  author  = {{Ripoche}, P. and {Heyl}, J. and {Parada}, J. and {Richer}, H.},
  title   = {{Carbon stars as standard candles: I. The luminosity function of
             carbon stars in the Magellanic Clouds}},
  journal = {\mnras}, year = 2020, volume = {495}, pages = {2858},
  doi     = {10.1093/mnras/staa1346}, eprint = {2005.05539}
}

@ARTICLE{Parada2021,
  author  = {{Parada}, J. and {Heyl}, J. and {Richer}, H. and {Ripoche}, P. and
             {Rousseau-Nepton}, L.},
  title   = {{Carbon stars as standard candles -- II. The median J magnitude as
             a distance indicator}},
  journal = {\mnras}, year = 2021, volume = {501}, pages = {933},
  doi     = {10.1093/mnras/staa3750}, eprint = {2011.11681}
}

@ARTICLE{Parada2023,
  author  = {{Parada}, J. and {Heyl}, J. and {Richer}, H. and {Ripoche}, P. and
             {Rousseau-Nepton}, L.},
  title   = {{Carbon stars as standard candles -- III. Un-binned maximum
             likelihood fitting and comparison with TRGB estimations}},
  journal = {\mnras}, year = 2023, volume = {522}, pages = {195},
  doi     = {10.1093/mnras/stad965}, eprint = {2303.16934}
}

@ARTICLE{Li2024,
  author  = {{Li}, S. and {Riess}, A.~G. and {Casertano}, S. and {Anand}, G.~S. and
             {Scolnic}, D.~M. and {Yuan}, W. and {Breuval}, L. and {Huang}, C.~D.},
  title   = {{Reconnaissance with JWST of the J-region Asymptotic Giant Branch in
             Distance Ladder Galaxies: From Irregular Luminosity Functions to
             Approximation of the Hubble Constant}},
  journal = {\apj}, year = 2024, volume = {966}, pages = {20},
  doi     = {10.3847/1538-4357/ad2f2b}, eprint = {2401.04777}
}

@ARTICLE{Li2025,
  author  = {{Li}, S. and {Riess}, A.~G. and {Scolnic}, D. and {Casertano}, S. and
             {Anand}, G.~S.},
  title   = {{JAGB 2.0: Improved Constraints on the J-region Asymptotic Giant
             Branch-based Hubble Constant from an Expanded Sample of JWST
             Observations}},
  journal = {\apj}, year = 2025, volume = {988}, pages = {97},
  doi     = {10.3847/1538-4357/addd0c}, eprint = {2502.05259}
}

@ARTICLE{Magnus2024,
  author  = {{Magnus}, E. and {Groenewegen}, M.~A.~T. and {Girardi}, L. and
             {Pastorelli}, G. and {Marigo}, P. and {Boyer}, M.~L.},
  title   = {{Calibration of the JAGB method for the Magellanic Clouds and Milky
             Way from Gaia DR3, considering the role of oxygen-rich AGB stars}},
  journal = {\aap}, year = 2024, volume = {691}, pages = {A350},
  doi     = {10.1051/0004-6361/202450677}, eprint = {2410.05974}
}

@ARTICLE{Lebzelter2018,
  author  = {{Lebzelter}, T. and {Mowlavi}, N. and {Marigo}, P. and
             {Pastorelli}, G. and {Trabucchi}, M. and {Wood}, P.~R. and
             {Lecoeur-Ta{\"\i}bi}, I.},
  title   = {{A new method to identify subclasses among AGB stars using
             {\it Gaia} and 2MASS photometry}},
  journal = {\aap}, year = 2018, volume = {616}, pages = {L13},
  doi     = {10.1051/0004-6361/201833615}, eprint = {1808.03659}
}

@ARTICLE{Lebzelter2023,
  author  = {{Lebzelter}, T. and {Mowlavi}, N. and {Lecoeur-Ta{\"\i}bi}, I. and
             {Trabucchi}, M. and {Audard}, M. and {Beck}, P.~G. and
             {Garc{\'\i}a-Lario}, P. and {Nienartowicz}, K. and {Rimoldini}, L.},
  title   = {{Gaia Data Release 3. The second {\it Gaia} catalogue of
             long-period variable candidates}},
  journal = {\aap}, year = 2023, volume = {674}, pages = {A15},
  doi     = {10.1051/0004-6361/202244241}, eprint = {2206.05745}
}

@INPROCEEDINGS{Mowlavi2019,
  author    = {{Mowlavi}, N. and {Trabucchi}, M. and {Lebzelter}, T.},
  title     = {{Long-period variables in the {\it Gaia} era}},
  booktitle = {The {\it Gaia} Universe},
  year      = 2019, eprint = {1907.05359}
}

@ARTICLE{Barnbaum1996,
  author  = {{Barnbaum}, C. and {Stone}, R.~P.~S. and {Keenan}, P.~C.},
  title   = {{A Moderate-Resolution Spectral Atlas of Carbon Stars: R, J, N, CH,
             and Barium Stars}},
  journal = {\apjs}, year = 1996, volume = {105}, pages = {419}
}

@ARTICLE{Keenan1954,
  author  = {{Keenan}, P.~C.},
  title   = {{Classification of the S-Type Stars}},
  journal = {\apj}, year = 1954, volume = {120}, pages = {484},
  doi     = {10.1086/145937}
}

@ARTICLE{KeenanBoeshaar1980,
  author  = {{Keenan}, P.~C. and {Boeshaar}, P.~C.},
  title   = {{Spectral types of S and SC stars on the revised MK system}},
  journal = {\apjs}, year = 1980, volume = {43}, pages = {379},
  doi     = {10.1086/190673}
}

@ARTICLE{Green2019,
  author  = {{Green}, G.~M. and {Schlafly}, E. and {Zucker}, C. and
             {Speagle}, J.~S. and {Finkbeiner}, D.},
  title   = {{A 3D Dust Map Based on Gaia, Pan-STARRS 1, and 2MASS}},
  journal = {\apj}, year = 2019, volume = {887}, pages = {93},
  doi     = {10.3847/1538-4357/ab5362}, eprint = {1905.02734}
}

@ARTICLE{BailerJones2021,
  author  = {{Bailer-Jones}, C.~A.~L. and {Rybizki}, J. and {Fouesneau}, M. and
             {Demleitner}, M. and {Andrae}, R.},
  title   = {{Estimating Distances from Parallaxes. V. Geometric and
             Photogeometric Distances to 1.47 Billion Stars in Gaia EDR3}},
  journal = {\aj}, year = 2021, volume = {161}, pages = {147},
  doi     = {10.3847/1538-3881/abd806}, eprint = {2012.05220}
}

@ARTICLE{GaiaDR3,
  author  = {{Gaia Collaboration} and {Vallenari}, A. and {Brown}, A.~G.~A. and
             {Prusti}, T. and others},
  title   = {{Gaia Data Release 3: Summary of the content and survey properties}},
  journal = {\aap}, year = 2023, volume = {674}, pages = {A1},
  doi     = {10.1051/0004-6361/202243940}
}

@ARTICLE{TwoMASS,
  author  = {{Skrutskie}, M.~F. and {Cutri}, R.~M. and {Stiening}, R. and others},
  title   = {{The Two Micron All Sky Survey (2MASS)}},
  journal = {\aj}, year = 2006, volume = {131}, pages = {1163},
  doi     = {10.1086/498708}
}

@ARTICLE{PypeIt,
  author  = {{Prochaska}, J.~X. and {Hennawi}, J.~F. and {Westfall}, K.~B. and
             {Cooke}, R.~J. and {Wang}, F. and {Hsyu}, T. and
             {Davies}, F.~B. and {Farina}, E.~P. and {Pelliccia}, D.},
  title   = {{PypeIt: The Python Spectroscopic Data Reduction Pipeline}},
  journal = {The Journal of Open Source Software}, year = 2020,
  volume  = {5}, pages = {2308}, doi = {10.21105/joss.02308}
}

@ARTICLE{SchlaflyFinkbeiner2011,
  author  = {{Schlafly}, E.~F. and {Finkbeiner}, D.~P.},
  title   = {{Measuring Reddening with Sloan Digital Sky Survey Stellar Spectra
             and Recalibrating SFD}},
  journal = {\apj}, year = 2011, volume = {737}, pages = {103},
  doi     = {10.1088/0004-637X/737/2/103}
}

@ARTICLE{SFD1998,
  author  = {{Schlegel}, D.~J. and {Finkbeiner}, D.~P. and {Davis}, M.},
  title   = {{Maps of Dust Infrared Emission for Use in Estimation of
             Reddening and Cosmic Microwave Background Radiation Foregrounds}},
  journal = {\apj}, year = 1998, volume = {500}, pages = {525},
  doi     = {10.1086/305772}, eprint = {astro-ph/9710327}
}

@ARTICLE{AbiaIsern2000,
  author  = {{Abia}, C. and {Isern}, J.},
  title   = {{The Chemical Composition of Carbon Stars. II. The J-Type Stars}},
  journal = {\apj}, year = 2000, volume = {536}, pages = {438},
  doi     = {10.1086/308932}
}

@ARTICLE{Filippenko1982,
  author  = {{Filippenko}, A.~V.},
  title   = {{The importance of atmospheric differential refraction in
             spectrophotometry}},
  journal = {\pasp}, year = 1982, volume = {94}, pages = {715},
  doi     = {10.1086/131052}
}

@ARTICLE{ChenYang2012,
  author  = {{Chen}, P.~S. and {Yang}, X.~H.},
  title   = {{A Catalog of Galactic Infrared Carbon Stars}},
  journal = {\aj}, year = 2012, volume = {143}, pages = {36},
  doi     = {10.1088/0004-6256/143/2/36}
}

@ARTICLE{Abia2022,
  author  = {{Abia}, C. and {de Laverny}, P. and {Romero-G{\'o}mez}, M. and
             {Figueras}, F.},
  title   = {{Characterisation of Galactic carbon stars and related stars
             from Gaia EDR3}},
  journal = {\aap}, year = 2022, volume = {664}, pages = {A45},
  doi     = {10.1051/0004-6361/202243595}, eprint = {2206.00405}
}

@ARTICLE{Li2024LAMOST,
  author  = {{Li}, L. and others},
  title   = {{Identification of Carbon Stars from LAMOST DR7}},
  journal = {\apjs}, year = 2024, volume = {271}, pages = {12},
  doi     = {10.3847/1538-4365/ad1881}
}

@ARTICLE{Creevey2023,
  author  = {{Gaia Collaboration} and {Creevey}, O.~L. and {Sarro}, L.~M. and
             {Lobel}, A. and {Pancino}, E. and {Andrae}, R. and others},
  title   = {{Gaia Data Release 3: A Golden Sample of Astrophysical
             Parameters}},
  journal = {\aap}, year = 2023, volume = {674}, pages = {A39},
  doi     = {10.1051/0004-6361/202243800}, eprint = {2206.05870}
}

@ARTICLE{Richer1981,
  author  = {{Richer}, H.~B.},
  title   = {{Observations of a Complete Sample of Carbon Stars in the Large
             Magellanic Cloud}},
  journal = {\apj}, year = 1981, volume = {243}, pages = {744}
}

@ARTICLE{Richer1984,
  author  = {{Richer}, H.~B. and {Crabtree}, D.~R. and {Pritchet}, C.~J.},
  title   = {{Luminous Late-Type Stars in NGC 205}},
  journal = {\apj}, year = 1984, volume = {287}, pages = {138}
}

@ARTICLE{Richer1985a,
  author  = {{Richer}, H.~B. and {Crabtree}, D.~R.},
  title   = {{Luminous Late-Type Stars in a Field of M31}},
  journal = {\apj}, year = 1985, volume = {298}, pages = {L13}
}

@ARTICLE{Richer1985b,
  author  = {{Richer}, H.~B. and {Pritchet}, C.~J. and {Crabtree}, D.~R.},
  title   = {{Luminous Late-Type Stars in NGC 300}},
  journal = {\apj}, year = 1985, volume = {298}, pages = {240}
}

@ARTICLE{Pritchet1987,
  author  = {{Pritchet}, C.~J. and {Richer}, H.~B. and {Schade}, D. and
             {Crabtree}, D. and {Yee}, H.~K.~C.},
  title   = {{The Late-Type Stellar Content of NGC 55}},
  journal = {\apj}, year = 1987, volume = {323}, pages = {79}
}

@ARTICLE{Cook1986,
  author  = {{Cook}, K.~H. and {Aaronson}, M. and {Norris}, J.},
  title   = {{Carbon and M Stars in Nearby Galaxies: A Preliminary Survey
             Using a Photometric Technique}},
  journal = {\apj}, year = 1986, volume = {305}, pages = {634}
}

@ARTICLE{Battinelli2005,
  author  = {{Battinelli}, P. and {Demers}, S.},
  title   = {{The Standard Candle Aspect of Carbon Stars}},
  journal = {\aap}, year = 2005, volume = {442}, pages = {159},
  doi     = {10.1051/0004-6361:20053357}
}

@ARTICLE{Madore2022,
  author  = {{Madore}, B.~F. and {Freedman}, W.~L. and {Lee}, A.~J. and
             {Owens}, K.},
  title   = {{Milky Way Zero-point Calibration of the JAGB Method: Using
             Thermally Pulsing AGB Stars in Galactic Open Clusters}},
  journal = {\apj}, year = 2022, volume = {938}, pages = {125},
  doi     = {10.3847/1538-4357/ac92fd}, eprint = {2209.08127}
}

@ARTICLE{LeeWLM2021,
  author  = {{Lee}, A.~J. and {Freedman}, W.~L. and {Madore}, B.~F. and
             {Owens}, K.~A. and {Monson}, A.~J. and {Hoyt}, T.~J.},
  title   = {{The Astrophysical Distance Scale. III. Distance to the Local
             Group Galaxy WLM Using Multiwavelength Observations of the Tip
             of the Red Giant Branch, Cepheids, and JAGB Stars}},
  journal = {\apj}, year = 2021, volume = {907}, pages = {112},
  eprint  = {2012.04536}
}

@ARTICLE{LeeM332022,
  author  = {{Lee}, A.~J. and {Rousseau-Nepton}, L. and {Freedman}, W.~L. and
             {Madore}, B.~F. and {Cioni}, M.-R.~L. and {Hoyt}, T.~J. and
             {Jang}, I.~S. and {Javadi}, A. and {Owens}, K.~A.},
  title   = {{The Astrophysical Distance Scale. V. A 2\% Distance to the
             Local Group Spiral M33 via the JAGB Method, Tip of the Red
             Giant Branch, and Leavitt Law}},
  journal = {\apj}, year = 2022, volume = {933}, pages = {201},
  doi     = {10.3847/1538-4357/ac7321}, eprint = {2205.11323}
}

@ARTICLE{Zgirski2021,
  author  = {{Zgirski}, B. and {Pietrzy{\'n}ski}, G. and {Gieren}, W. and
             {G{\'o}rski}, M. and {Wielg{\'o}rski}, P. and {Karczmarek}, P. and
             {Bresolin}, F. and {Kervella}, P. and {Kudritzki}, R.-P. and
             {Storm}, J. and {Graczyk}, D. and {Hajdu}, G. and {Narloch}, W. and
             {Pilecki}, B. and {Suchomska}, K. and {Taormina}, M.},
  title   = {{The Araucaria Project. Distances to Nine Galaxies Based on a
             Statistical Analysis of their Carbon Stars (JAGB Method)}},
  journal = {\apj}, year = 2021, volume = {916}, pages = {19},
  eprint  = {2105.02120}
}

@ARTICLE{LeeMagellan2024,
  author  = {{Lee}, A.~J. and {Monson}, A.~J. and {Freedman}, W.~L. and
             {Madore}, B.~F. and {Owens}, K.~A. and {Beaton}, R.~L. and
             {Espinoza}, C. and {Ren}, T. and {Ren}, Y.},
  title   = {{Resolved Near-infrared Stellar Photometry from the Magellan
             Telescope for 13 Nearby Galaxies: JAGB Method Distances}},
  journal = {\apj}, year = 2024, volume = {967}, pages = {22},
  doi     = {10.3847/1538-4357/ad32c7}, eprint = {2402.18794}
}

@ARTICLE{LeeJWST2024,
  author  = {{Lee}, A.~J. and {Freedman}, W.~L. and {Jang}, I.~S. and
             {Madore}, B.~F. and {Owens}, K.~A.},
  title   = {{First JWST Observations of JAGB Stars in the SN Ia Host
             Galaxies: NGC 7250, NGC 4536, NGC 3972}},
  journal = {\apj}, year = 2024, volume = {961}, pages = {132},
  eprint  = {2312.02282}
}

@ARTICLE{LeeJWSTH02024,
  author  = {{Lee}, A.~J. and {Freedman}, W.~L. and {Madore}, B.~F. and
             {Jang}, I.~S. and {Owens}, K.~A. and {Hoyt}, T.~J.},
  title   = {{The Chicago-Carnegie Hubble Program: The JWST J-region
             Asymptotic Giant Branch (JAGB) Extragalactic Distance Scale}},
  journal = {arXiv e-prints}, year = 2024, eprint = {2408.03474}
}

@ARTICLE{Aringer2016,
  author  = {{Aringer}, B. and {Girardi}, L. and {Nowotny}, W. and
             {Marigo}, P. and {Bressan}, A.},
  title   = {{Synthetic photometry for M and K giants and stellar evolution:
             hydrostatic dust-free model atmospheres and chemical
             abundances}},
  journal = {\mnras}, year = 2016, volume = {457}, pages = {3611},
  doi     = {10.1093/mnras/stw222}
}

@ARTICLE{Aringer2019,
  author  = {{Aringer}, B. and {Marigo}, P. and {Nowotny}, W. and
             {Girardi}, L. and {Me{\v{c}}ina}, M. and {Nanni}, A.},
  title   = {{Carbon stars with increased oxygen and nitrogen abundances:
             hydrostatic dust-free model atmospheres}},
  journal = {\mnras}, year = 2019, volume = {487}, pages = {2133},
  doi     = {10.1093/mnras/stz1429}
}

@ARTICLE{Barber2006,
       author = {{Barber}, R.~J. and {Tennyson}, J. and {Harris}, G.~J. and {Tolchenov}, R.~N.},
        title = "{A high-accuracy computed water line list}",
      journal = {\mnras},
         year = 2006,
        month = may,
       volume = {368},
       number = {3},
        pages = {1087-1094},
          doi = {10.1111/j.1365-2966.2006.10184.x},
archivePrefix = {arXiv},
       eprint = {astro-ph/0601236},
 primaryClass = {astro-ph},
       adsurl = {https://ui.adsabs.harvard.edu/abs/2006MNRAS.368.1087B}
}

@ARTICLE{Sanchez-Blazquez2006,
  author  = {{Sanchez-Blazquez}, P. and {Peletier}, R.~F. and
             {Jimenez-Vicente}, J. and {Cardiel}, N. and {Cenarro}, A.~J. and
             {Falcon-Barroso}, J. and {Gorgas}, J. and {Selam}, S. and
             {Vazdekis}, A.},
  title   = {{MILES: A Medium resolution INT Library of Empirical Spectra}},
  journal = {\mnras}, year = 2006, volume = {371}, pages = {703},
  doi     = {10.1111/j.1365-2966.2006.10699.x}
}

@ARTICLE{Steinmetz2006,
  author  = {{Steinmetz}, M. and {Zwitter}, T. and {Siebert}, A. and others},
  title   = {{The Radial Velocity Experiment (RAVE): First Data Release}},
  journal = {\aj}, year = 2006, volume = {132}, pages = {1645},
  doi     = {10.1086/506564}
}

@ARTICLE{DeSilva2015,
  author  = {{De Silva}, G.~M. and {Freeman}, K.~C. and
             {Bland-Hawthorn}, J. and others},
  title   = {{The GALAH survey: scientific motivation}},
  journal = {\mnras}, year = 2015, volume = {449}, pages = {2604},
  doi     = {10.1093/mnras/stv327}
}

@ARTICLE{York2000,
  author  = {{York}, D.~G. and {Adelman}, J. and {Anderson}, John~E., Jr. and
             others},
  title   = {{The Sloan Digital Sky Survey: Technical Summary}},
  journal = {\aj}, year = 2000, volume = {120}, pages = {1579},
  doi     = {10.1086/301513}
}

@ARTICLE{Cui2012,
  author  = {{Cui}, Xiang-Qun and {Zhao}, Yong-Heng and {Chu}, Yao-Quan and
             others},
  title   = {{The Large Sky Area Multi-Object Fiber Spectroscopic Telescope
             (LAMOST)}},
  journal = {\raa}, year = 2012, volume = {12}, pages = {1197},
  doi     = {10.1088/1674-4527/12/9/003}
}

@ARTICLE{LopezMarti2025,
  author  = {{L{\'o}pez Mart{\'\i}}, B. and {Jim{\'e}nez-Esteban}, F.~M. and
             {Engels}, D. and {Garc{\'\i}a-Lario}, P.},
  title   = {{The Gaia Catalogue of Galactic AGB Stars. I. OH/IR stars}},
  journal = {\aap}, year = 2025, volume = {698}, pages = {A109},
  doi     = {10.1051/0004-6361/202453125}
}

@ARTICLE{Stephenson1984,
  author  = {{Stephenson}, C.~B.},
  title   = {{A general catalogue of galactic S stars, 2nd edition}},
  journal = {Publications of the Warner and Swasey Observatory},
  year    = 1984, volume = {3}, pages = {1}
}

@ARTICLE{KenyonFernandezCastro1987,
  author  = {{Kenyon}, S.~J. and {Fernandez-Castro}, T.},
  title   = {{The cool components of symbiotic stars. I - Optical spectral types}},
  journal = {\aj}, year = 1987, volume = {93}, pages = {938}
}

\end{document}